\documentclass[twocolumn]{aastex701}

\received{XXX}
\revised{YYY}
\accepted{ZZZ}

\usepackage{amsmath}
\usepackage{enumitem} % Required for customizing enumerate lists
\usepackage{natbib}

\defcitealias{McVenn2020}{MV20a}
\defcitealias{McVenn2020_updated}{MV20b}
\defcitealias{thomas2018}{T18}
\defcitealias{thomas2019}{T19}

\begin{document}

% \title{Connecting the dots: linking the stellar debris of Bo\"{o}tes~3 to the Styx stellar stream}
\title{Characterizing the disruption of Bo\"otes~III: a missing link in the Galactic halo?}
\shorttitle{Characterizing the disruption of Bo\"otes~III}
\shortauthors{J. Jensen et al.}

\author[0000-0002-4350-7632]{Jaclyn Jensen}
\email[show]{jaclyn.r.jensen@dartmouth.edu}
\affiliation{Department of Physics \& Astronomy, Dartmouth College, 6127 Wilder Laboratory, Hanover, NH 03755, USA}
\affiliation{Department of Physics \& Astronomy, University of Victoria, Victoria, BC, V8P 1A1, Canada}
% \affiliation{NRC Herzberg Astronomy \& Astrophysics, 5071 West Saanich Road, Victoria, BC, V9E 2E7, Canada}

\author[0009-0008-8903-160X]{Daniel Boyea}
\email{danielaboyea@gmail.com}
\affiliation{Department of Physics \& Astronomy, University of Victoria, Victoria, BC, V8P 1A1, Canada}

\author[0000-0003-4666-6564]{Alan W. McConnachie}
\email{alan.mcconnachie@nrc-cnrc.gc.ca}
\affiliation{NRC Herzberg Astronomy \& Astrophysics, 5071 West Saanich Road, Victoria, BC, V9E 2E7, Canada}
\affiliation{Department of Physics \& Astronomy, University of Victoria, Victoria, BC, V8P 1A1, Canada}

\author{Rapha\"el Errani}
\email{errani@cmu.edu}
\affiliation{McWilliams Center for Cosmology, Department of Physics, Carnegie Mellon University, Pittsburgh, PA 15213, USA}

\author[0000-0002-7507-5985]{Akshara Viswanathan}
\email{aksharaviswanathan@uvic.ca}
\affiliation{Department of Physics \& Astronomy, University of Victoria, Victoria, BC, V8P 1A1, Canada}

% \collaboration{20}{(Pristine Collaboration)}

\author[0000-0002-1349-202X]{Nicolas Martin}
\email{nicolas.martin@astro.unistra.fr}
\affiliation{l’Universit\'{e} de Strasbourg, CNRS, Observatoire astronomique de Strasbourg, UMR 7550, F-67000 Strasbourg, France}
\affiliation{Max-Planck-Institut f\"{u}r Astronomie, K\"{o}nigstuhl 17, D-69117 Heidelberg, Germany}

\author[0000-0002-8129-5415]{Zhen Yuan}
\email{zhen.yuan@nju.edu.cn}
\affiliation{l’Universit\'{e} de Strasbourg, CNRS, Observatoire astronomique de Strasbourg, UMR 7550, F-67000 Strasbourg, France}
\affiliation{Key Laboratory of Modern Astronomy and Astrophysics, Nanjing University, Ministry of Education, Nanjing 210093, China}

% \author[0000-0002-0544-2217]{Anke Ardern-Arentsen}
% \affiliation{Institute of Astronomy, University of Cambridge, Madingley Road, Cambridge CB3 0HA, UK}

\author[0000-0002-2468-5521]{Guillaume F. Thomas}
\email{guillaume.thomas@iac.es}
\affiliation{Universidad de La Laguna, Dpto. Astrofísica, E-38206 La Laguna, Tenerife, Spain}
\affiliation{Instituto de Astrofísica de Canarias, E-38205 La Laguna, Tenerife, Spain}

\author[0000-0003-3862-5076]{Julio F. Navarro}
\email{jfn@uvic.ca}
\affiliation{Department of Physics \& Astronomy, University of Victoria, Victoria, BC, V8P 1A1, Canada}
\affiliation{Crimson Distinguished Professor, KU-KIST Green School, Korea University}

\author[0000-0003-4134-2042]{Kim Venn}
\email{kvenn@uvic.ca}
\affiliation{Department of Physics \& Astronomy, University of Victoria, Victoria, BC, V8P 1A1, Canada}

\author[0000-0002-9655-1063]{Pascale Jablonka}
\email{pascale.jablonka@epfl.ch}
\affiliation{Institute of Physics, Laboratory of Astrophysics, \'{E}cole Polytechnique F\'{e}d\'{e}rale de Lausanne (EPFL), Switzerland}

\author[0000-0002-8318-433X]{Khyati Malhan}
\email{kmalhan07@gmail.com}
\affiliation{Independent Researcher}

\author[0009-0001-1871-2001]{Anya Dovgal}
\email{adovgal@uvic.ca}
\affiliation{Department of Physics \& Astronomy, University of Victoria, Victoria, BC, V8P 1A1, Canada}

\author[0000-0002-6946-8280]{Simon E.T. Smith}
\email{simonsmith@uvic.ca}
\affiliation{Department of Physics \& Astronomy, University of Victoria, Victoria, BC, V8P 1A1, Canada}
% \affiliation{NRC Herzberg Astronomy \& Astrophysics, 5071 West Saanich Road, Victoria, BC, V9E 2E7, Canada}

\author[0000-0003-0105-9576]{Gustavo~E.~Medina}
\email{gustavo.medina@utoronto.ca}
\affiliation{David A. Dunlap Department of Astronomy \& Astrophysics, University of Toronto, 50 St George Street, Toronto ON M5S 3H4, Canada}
\affiliation{Dunlap Institute for Astronomy \& Astrophysics, University of Toronto, 50 St George Street, Toronto, ON M5S 3H4, Canada}

%%%%UNIONS steering group
\author[0000-0001-5486-2747]{Thomas de Boer}
\email{tdeboer@hawaii.edu}
\affiliation{Institute for Astronomy, University of Hawai`i, 2680 Woodlawn Drive, Honolulu, HI 96822, USA}

\author[0000-0002-6639-6533,gname='Gregory', sname='Paek']{Gregory S. H. Paek}  
\email{gregorypaek94@gmail.com}
\affiliation{Institute for Astronomy, University of Hawai`i, 2680 Woodlawn Drive, Honolulu, HI 96822, USA}

% \collaboration{20}{(\& UNIONS Collaboration)}

% \author[]{Guillaume Thomas ?}
% \affiliation{}

% \author[]{}
% \affiliation{}

\begin{abstract}

% . This makes a putative link to Styx seem very reasonable, but as yet a clear connection between the two systems has not been demonstrated.

The Bo\"otes III (Boo3) dwarf galaxy has long been suspected of being the progenitor of Styx, a $\sim$50$^{\circ}$-long stellar stream that was simultaneously discovered in the same region of sky. Boo3’s diffuse morphology, large velocity dispersion, small pericenter, and excess of candidate stars at large radii suggest it is undergoing active tidal disruption. A link to Styx is therefore logical; however, a clear connection between these structures has not yet been clearly demonstrated. Here, we re-examine the Boo3-Styx association by searching for Boo3’s tidal debris using a combination of \textit{Gaia}-selected members, new CaHK narrow-band imaging with CFHT/MegaCam, and stellar tracer catalogues of blue horizontal branch and red giant branch stars. We also conduct a broad search for a putative stream using matched filter techniques applied to SDSS DR17 and DELVE DR2. Despite our extensive search, we find no observational evidence directly linking Boo3 to Styx. Furthermore, our results suggest that either Boo3’s extended substructure is too diffuse to be detected with current data, or that its particular orbit may have erased a coherent tidal signature. Boo3 thus remains an enigmatic system, and exemplifies the need for spectroscopic follow-up to properly disentangle the nature between this faint Milky Way satellite and nearby stream. 
% or that a more complex dynamical process (such as interaction with the Galactic bar) has erased a coherent tidal signature
% to disentangle the nature of faint Milky Way substructure. 

% This example manuscript is intended to serve as a tutorial and template for
% authors to use when writing their own AAS Journal articles. The manuscript
% includes a history of \aastex\ and includes figure and table examples to illustrate these features. Information on features not explicitly mentioned in the article can be viewed in the manuscript comments or more extensive online
% documentation. Authors are welcome replace the text, tables, figures, and
% bibliography with their own and submit the resulting manuscript to the AAS
% Journals peer review system.  The first lesson in the tutorial is to remind
% authors that the AAS Journals, the Astrophysical Journal (ApJ), the
% Astrophysical Journal Letters (ApJL), the Astronomical Journal (AJ), and
% the Planetary Science Journal (PSJ) all have a 250 word limit for the 
% abstract\footnote{Abstracts for Research Notes of the American Astronomical 
% Society (RNAAS) are limited to 150 words}.  If you exceed this length the
% Editorial office will ask you to shorten it. This abstract has 161 words.

\end{abstract}

%% Keywords should appear after the \end{abstract} command. 
%% The AAS Journals now uses Unified Astronomy Thesaurus concepts:
%% https://astrothesaurus.org
%% You will be asked to selected these concepts during the submission process
%% but this old "keyword" functionality is maintained in case authors want
%% to include these concepts in their preprints.
% \keywords{\uat{Dwarf galaxies}{416}}

%% From the front matter, we move on to the body of the paper.
%% Sections are demarcated by \section and \subsection, respectively.
%% Observe the use of the LaTeX \label
%% command after the \subsection to give a symbolic KEY to the
%% subsection for cross-referencing in a \ref command.
%% You can use LaTeX's \ref and \label commands to keep track of
%% cross-references to sections, equations, tables, and figures.
%% That way, if you change the order of any elements, LaTeX will
%% automatically renumber them.
%%
%% We recommend that authors also use the natbib \citep
%% and \citet commands to identify citations.  The citations are
%% tied to the reference list via symbolic KEYs. The KEY corresponds
%% to the KEY in the \bibitem in the reference list below. 

%%%%%%%%%%%%%%%%%%%%%%%%%%%%%%%%%%%%%%%%%%%%%%%%%%%%%%%%%%%%%%
\section{Introduction} 
\label{section:intro}

In the prevailing $\Lambda$CDM cosmological model, galaxies like the Milky Way (MW) are assembled hierarchically through the accretion and merging of smaller systems (\citealt{white_rees1978, frenk1988}). Dwarfs $-$ the most numerous type of galaxy in the Universe $-$ are considered the fundamental building blocks of larger galaxies as their stars and dark matter directly contribute to the haloes of their host (\citealt{bullock2005, cooper2010}). As these low-mass satellites fall into the gravitational potential of a more massive galaxy like the MW, they experience tidal forces imparted by the host that gradually strips away their mass. Over time, this process results in the decimation of a satellite, whose matter becomes strewn across the host in the form of lengthy tidal tails until the tidally stripped material becomes phase-mixed and assimilated into the more massive galaxy.

The stellar halo of our own Galaxy is similarly populated by substructures in various stages of disruption (\citealt{shipp2024}), ranging from intact satellites (dwarf galaxies and globular clusters; \citealt{mcconnachie2012, harris1996}), to partially disrupted systems (stellar streams, with and without clear progenitors; e.g., \citealt{malhan2018, ibata2024, mateu2023}), and fully disrupted phase-mixed debris that may only retain chemical or kinematic signatures of their origin (e.g., \citealt{helmi1999, helmi2018, naidu2020, horta2023, dodd2025, malhanrix2024}). Studying these structures not only reveals our own Galaxy's accretion history and the underlying gravitational potential of the MW (\citealt{johnston1999, bullock2005}), but can also inform on the resilience of accreted dwarf galaxies and of their own dark matter (DM) haloes (\citealt{penarrubia2008, penarrubia2010, errani2020}), thereby offering insight into the nature of DM and small-scale galaxy formation (\citealt{bullock_boylan-kolchin2017, sales2022}).

% (see \citealt{li2022})

Bo\"otes 3 (Boo3) is one such dwarf galaxy that may be undergoing the final stages of tidal disruption. Photometric observations indicate that Boo3 is particularly diffuse, large (half-light radius of 33.03$\arcmin$, or $\sim$450~pc; \citealt{moskowitz2020}), and morphologically, appears double-lobed and misshapen  (see Figure 10 in \citealt{grillmair2009}), which may indicate that the system is not currently dynamically relaxed. Simultaneous to the discovery of Boo3, \citet{grillmair2009} identified a $\sim$50$^{\circ}$-long and spatially-coincident stellar stream, known as Styx. These initial observations suggested that Boo3 is the progenitor (or at minimum, associated with) Styx. However, both of these structures have remained largely unstudied since their original detection, in part due to the faint (M$_{V}$ = $-$5.75 mag; \citealt{correnti2009}) and diffuse nature ($\mu$ = 31 mag arcmin$^{-2}$; \citealt{pace2024}) of Boo3, making follow-up observations challenging. 
% Boo3's on-sky position also lies coincident with a $\sim$50$^{\circ}$-long structure, known as the Styx stellar stream, that was simultaneously discovered with Boo3 by \citet{grillmair2009}. 

Previous work by \citet{jensen2024} indicated that Boo3 is also one of a handful of MW dwarf galaxies that exhibit evidence of an extended stellar distribution, where a clear excess in the density profile (above what is expected for a single component system) was identified. Detected in photometric and astrometric data from the third \textit{Gaia} (early) data release (eDR3; \citealt{gaia_edr3_2021}), \citet{jensen2024} reported that Boo3's stellar excess extends well beyond $\sim$3 half-light radii ($r_{h}$), and in fact may reach as far as $\sim$14$r_{h}$. One possible interpretation is that this outer component, particularly in lower-mass dwarfs (Ultra-Faint Dwarfs, or UFDs) such as Boo3, originates because of external factors (i.e., MW tides inducing tidal stripping such that \textit{in situ} stars migrate outwards, or due to a previously accreted system whose stars then form the host dwarf's stellar halo). As our own previous work indicates that there is an extended stellar population in the outskirts of Boo3, we aim to explore here whether this feature is tidal in origin.

Despite the lack of recent studies, a number of previous results indeed suggest that Boo3 is tidally disrupting. Recent constraints of Boo3’s kinematics, such as proper motion (\citealt{carlin2018, pace2022, battaglia2022}) and radial velocity (\citealt{carlin2009}), suggest a very small pericenter of only 7 $-$ 9 kpc (\citealt{pace2022, battaglia2022}) which implies there may be strong tidal influence from the MW on this particular dwarf. This conclusion is also in agreement with Boo3's large velocity dispersion ($\sigma_{\rm RV}$~$\sim$~14~km~s$^{-1}$; \citealt{carlin2009}) which may have been inflated substantially as the system has lost mass (e.g., see the dissolution of Ursa Major~2 in \citealt{fellhauer2007}).

% where the velocity dispersion is found to increase by a factor of 10 before dissolution).
% see simulations of Ursa Major~2

% (\citealt{errani2021, errani2022}). \textcolor{red}{Address JFN and RE comment RE: inflation of vel disp}

% , which is nearly as large as Sculptor, Ursa Minor, and other (much) more luminous MW dwarfs. 

Though there is substantial evidence to support that Boo3 is being tidally disrupted, it remains true that (i) no spectroscopically-confirmed members of Boo3's stellar debris/stream have yet been catalogued or reported in the literature\footnote{While in the process of submitting this manuscript, we became aware of a recent report from \citet{li2026} in which 6 stars were spectroscopically-confirmed to be members of Boo3, located between 3 $-$ 6$r_{h}$.} and (ii) a link between Boo3 and Styx has not been officially confirmed. In this work, we closely inspect the Boo3 dwarf with an intent to detect its putative tidal tails. If Boo3 is shown to possess a stream, it will join the exceedingly small sample of \textit{surviving} dwarf galaxy progenitors that are reported to have lengthy (i.e., $>$ multiple degree-long) tidal tails: Sagittarius (\citealt{ibata2002, majewski2003}) and Crater~2 (\citealt{limberg2025, atzberger2026}). 
% Tucana~3 (\citealt{drlica-wagner2015, li2018}), 
% Indeed, This putative verification could yield interesting results for future dynamical studies exploring the underlying mass and shape of our Galaxy’s dark matter halo. 

% If instead we are unable to resolve Boo3’s tidal debris, it may be reasonable to conclude that (i) Boo3 and Styx are unrelated structures, if indeed Styx can be re-detected, (ii) Boo3’s tidal tails may have been dispersed during an interaction, such as with the Galactic bar, or (iii) Boo3 may be the MW’s first ultra-diffuse dwarf galaxy (UDG) whose retained structure indicates the presence of a significant dark matter halo \textcolor{red}{SOURCE?}. 

We organize this paper as follows. Section \ref{section:dynamics} details the initial conditions of the dynamical models (a point-mass orbit and particle-spray simulation) used in this work to inform on the likely phase-space positions of Boo3's stellar debris. We then present in Section \ref{section:data} the various datasets (\textit{Gaia} members from \citealt{jensen2024}, follow-up CaHK observations, stellar tracer catalogues from the Ultraviolet Near-Infrared Optical Northern Survey, and wide-field photometric surveys) used in this work, as well as our calibrations to these data. The derivations of photometric metallicity from the CaHK follow-up observations are found in Section \ref{section:inner}. Section \ref{section:outer} presents the analysis and results of our search for tidal debris. We discuss the observational and dynamical results in Section \ref{section:disc} and conclude in Section \ref{section:summary}. 
% in our stellar tracer catalogues and the photometric surveys

% 

\begingroup
    \renewcommand{\arraystretch}{1.1} % Default value: 1; vertical spacing
    \begin{table}
        \centering
        \begin{tabular}{ccc}
            \hline
            Observables & Value & Refs. \\
            \hline
            ($\alpha$, $\delta$)              & (209.30$^{\circ}$, 26.80$^{\circ}$)            & (1) \\
            R$_{helio}$                       & 46.56 $\pm$ 4 kpc                              & (2) \\
            $r_{h}$                           & 33.03 $\pm$ 2.50 arcmin                        & (3) \\
                                              & 447 $\pm$ 34 pc                                & (3) \\
            PA                                & $-$81$^{\circ}$ $\pm$ 8$^{\circ}$            & (3) \\
            $\epsilon$                        & 0.33 $\pm$ 0.09                                & (3) \\
            M$_{V}$                           & $-$5.74 $\pm$ 0.5 mag                             & (5) \\
            $\rm [Fe/H]$$^{\dagger}$              & $-$2.1 $\pm$ 0.2                               & (6) \\
                                              & $-$2.1 $\pm$ 0.1                               & This work \\
            $\sigma_{\rm Fe/H}$$^{\dagger}$       & 0.6 $\pm$ 0.2                                  & (6) \\
                                              & 0.2 $\pm$ 0.1                                  & This work \\
            $\mu_{\alpha*}$                   & $-$1.160 $\pm$ 0.037 mas~yr$^{-1}$             & (4) \\
            $\mu_{\delta}$                    & $-$0.880 $\pm$ 0.035 mas~yr$^{-1}$             & (4) \\
            RV                                & 197.5 $\pm$ 3.8 km~s$^{-1}$                    & ($-$) \\
            $\sigma_{\rm RV}$                     & 14.0 $\pm$ 3.2 km~s$^{-1}$                     & (6) \\
            
           \hline
            
        \end{tabular}
        \caption{Observational properties of Boo3 relevant to this work. From top to bottom, the rows indicate Boo3's equatorial coordinates (RA, Dec), heliocentric distance, half-light radius, position angle, ellipticity, absolute magnitude, metallicity, metallicity dispersion, proper motion in RA and Dec, radial velocity, and radial velocity dispersion. $^{\dagger}$Observables with new measurements derived in this work. Citations: (1) \citet{grillmair2009}, (2) \citet{vivas2020}, (3) \citet{moskowitz2020}, (4) \citet{jensen2024}, (5) \citet{correnti2009}, (6) \citet{carlin2009}.}
        \label{tab:Boo3Params}
    \end{table}
\endgroup
% (3) \citet{carlin2018},

%%%%%%%%%%%%%%%%%%%%%%%%%%%%%%%%%%%%%%%%%%%%%%%%%%%%%%%%%%%%%%
\section{Modeling the Orbit of Bo\"otes~III}
% \section{Outer Study of Bo\"otes 3: a Search for Styx}
\label{section:dynamics}

To help guide our search for a coherent tidal structure from Boo3, we first compute both the satellite's orbit and the expected positions and velocities of potential debris. We emphasize that we do not rely on these models to \textit{select} candidate stream stars, but instead use them to independently compare expected trends of \textit{bona fide} stream members. In this section, we first present the initial conditions and adopted parameters for these models. Then, we summarize our dynamical methods, for which we adopt two complementary approaches. These involve (i) determining Boo3’s point-mass orbit to characterize the trajectory of the satellite, and (ii) using a ``particle-spray'' technique (\citealt{fardal2015}) where massless particles are ``ejected'' at specified timesteps along the system's orbit and integrated forward in time, allowing us to illustrate the phase-space distribution of putative debris. In both methods, we examine the dynamical differences when assuming a static versus a time-evolving (i.e., one that included the influence of a massive LMC) potential. Note that Boo3's point-mass orbit will appear later in Sections~\ref{section:BHB_RGBs} and \ref{section:outer_MF}, and that we reserve a thorough discussion of Boo3's dynamics for Section~\ref{section:disc}.

\subsection{Initial Conditions}
\label{section:dynm_init}

In this work, we calculate Boo3's orbit and create a particle-spray model of its debris using the \texttt{AGAMA} framework (\citealt{vasiliev2018_agama}). Initial conditions for Boo3 are determined using the present-day observations of the system, summarized in Table \ref{tab:Boo3Params}. Conversion between Galactic and Equatorial coordinate frames are conducted using \texttt{astropy} (\citealt{astropy2013, astropy2018}), where we assume a right-handed Galactocentric cartesian coordinate system (X,~Y,~Z). In this frame, the Sun is located at ($-$8.122,~0.000,~0.025) kpc (\citealt{gravity2019}; \citealt{juric2008}) with a circular velocity of 229~km~s$^{-1}$ (\citealt{eilers2019}) and we assume the Solar peculiar motion is [U,~V,~W]$_{\odot}$~=~[11.1,~12.24,~7.25]~km~s$^{-1}$ (\citealt{schonrich2010}).

To explore the impact of the LMC in Boo3's trajectory, we use both a static and a time-evolving potential. Both MW potentials consist of three main components: a spherical bulge, an exponential disk, and a triaxial dark matter halo. The static potential, modeled after that of \citet{thomas2022}, is a composite whose parameters (scale lengths, densities, and axis ratios) have been tuned to best replicate the observed MW rotation curve (\citealt{eilers2019}). For the evolving potential, we utilize the pre-set triaxial model developed by \citet{vasiliev2021_tango} which has been calibrated to match the MW rotation curve and provides an excellent fit to stars in the Sagittarius stream. The evolving potential includes the orbit and gravitational wake of a massive LMC (1.5~$\times$~10$^{11}$~$M_{\odot}$) which induces variations in the MW potential over time. The LMC in this model is on first infall, entering the MW halo only 1.6 Gyrs ago.

\begin{figure}
    \centering
    \includegraphics[width=\linewidth]{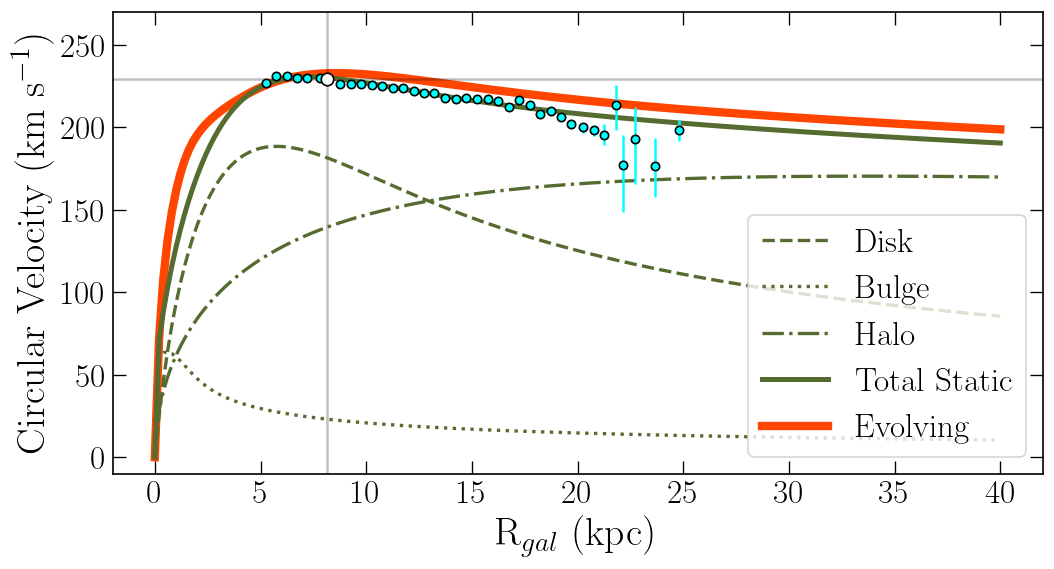}
    \caption{Rotation curves for the total static potential (consisting of a composite 3-component disk, a bulge, and a DM halo) in olive and a time-evolving potential (including the LMC) in orange. We find good agreement between our chosen models in comparison to the MW's observed rotation curve (\citealt{eilers2019}; cyan error bars), and also note that both models accurately reproduce the circular velocity of the Sun at v$_{circ}$(R$_{\odot}$) = 229.0 km s$^{-1}$ (grey cross-hairs).}
    \label{fig:rot_curve}
\end{figure}

\renewcommand{\arraystretch}{1.5}
\begin{table*}
    \centering
    \setlength{\tabcolsep}{4pt}
    \begin{tabular}{c||cc|cc}
    \hline
         Component & Static & Evolving & Observations & Refs.  \\
                   % & (M$_{\odot}$) & (M$_{\odot}$) & (M$_{\odot}$) &   \\
         \hline
         Disk (M$_{\odot}$)           &  6.5~$\times$~10$^{10}$    &   5.0~$\times$~10$^{10}$      & (4.6 $-$ 5.7)~$\times$~10$^{10}$               & (1); (2); (3) \\
         Bulge (M$_{\odot}$)          &  1.0~$\times$~10$^{9}$     &   1.2~$\times$~10$^{10}$      & (0.9 $-$ 2.2)~$\times$~10$^{10}$               & (2); (4); (5) \\
         Halo ($M_{200}$ M$_{\odot}$)           &  0.7~$\times$~10$^{12}$   &   0.88~$\times$~10$^{12}$     & (0.8 $-$ 1.3)~$\times$~10$^{12}$  & (6); (7); (8) \\
         \hline
         LMC ($M_{200}$ M$_{\odot}$)             &    $-$                                &   1.5~$\times$~10$^{11}$      & (1.3 $-$ 1.9)~$\times$~10$^{11}$               & (9); (10); (11) \\
         \hline
    \end{tabular}
    \caption{Enclosed masses for MW components used in the static and evolving potential within 300 kpc of Galactic center, compared to current literature estimates. Citations: (1) \citet{bovy_rix2013}, (2) \citet{licquia_newman2015}, (3) \citet{mcmillan2017}, (4) \citet{valenti2016}, (5) \citet{portail2017}, (6) \citet{callingham2019}, (7) \citet{cautun2020}, (8) \citet{deason2021}, (9) \citet{erkal2019}, (10) \citet{vasiliev2021_tango}, (11) \citet{shipp2021}.}
    \label{tab:potential_params}
\end{table*}

% As demonstrated in Figure~\ref{fig:rot_curve}, both potentials are suitable for our use. These rotation curves show that both the 

Figure \ref{fig:rot_curve} shows that the circular velocity rotation curves of both the static (green) and evolving (orange) potentials are consistent with the observed MW rotation curve (cyan error bars) and match the circular velocity of the Sun (grey cross-hairs) reasonably well. Although these potentials differ in construction, the enclosed masses of each MW component are broadly consistent to observed estimates. We provide these values in Table~\ref{tab:potential_params} for reference, and include the ranges of reported literature values for comparison.

\subsection{Satellite Orbit \& Phase-Space of Stream}
\label{section:orb_partspray}

We calculate the orbit of Boo3, represented by a point-mass, in both the static and evolving potentials. In the pre-set model from \citet{vasiliev2021_tango}, the MW's potential over each timestep is represented by snapshots that were taken from a more detailed \texttt{AGAMA} N-body simulation. While using this version of the evolving potential is more accessible and takes less computation time than a full N-body simulation, the \citet{vasiliev2021_tango} evolving potential does not currently include the orbit of the LMC in timesteps past the present day. As such, we only use this model to integrate the orbit of Boo3 backwards in time by 5 Gyrs, in timesteps of 5 Myrs. As the static potential does not vary with time, we integrate it both forwards and backwards in time (i.e., $\pm$5 Gyrs) in timesteps of 5 Myrs. 

\begin{figure}
    \centering

    \includegraphics[width=\linewidth]{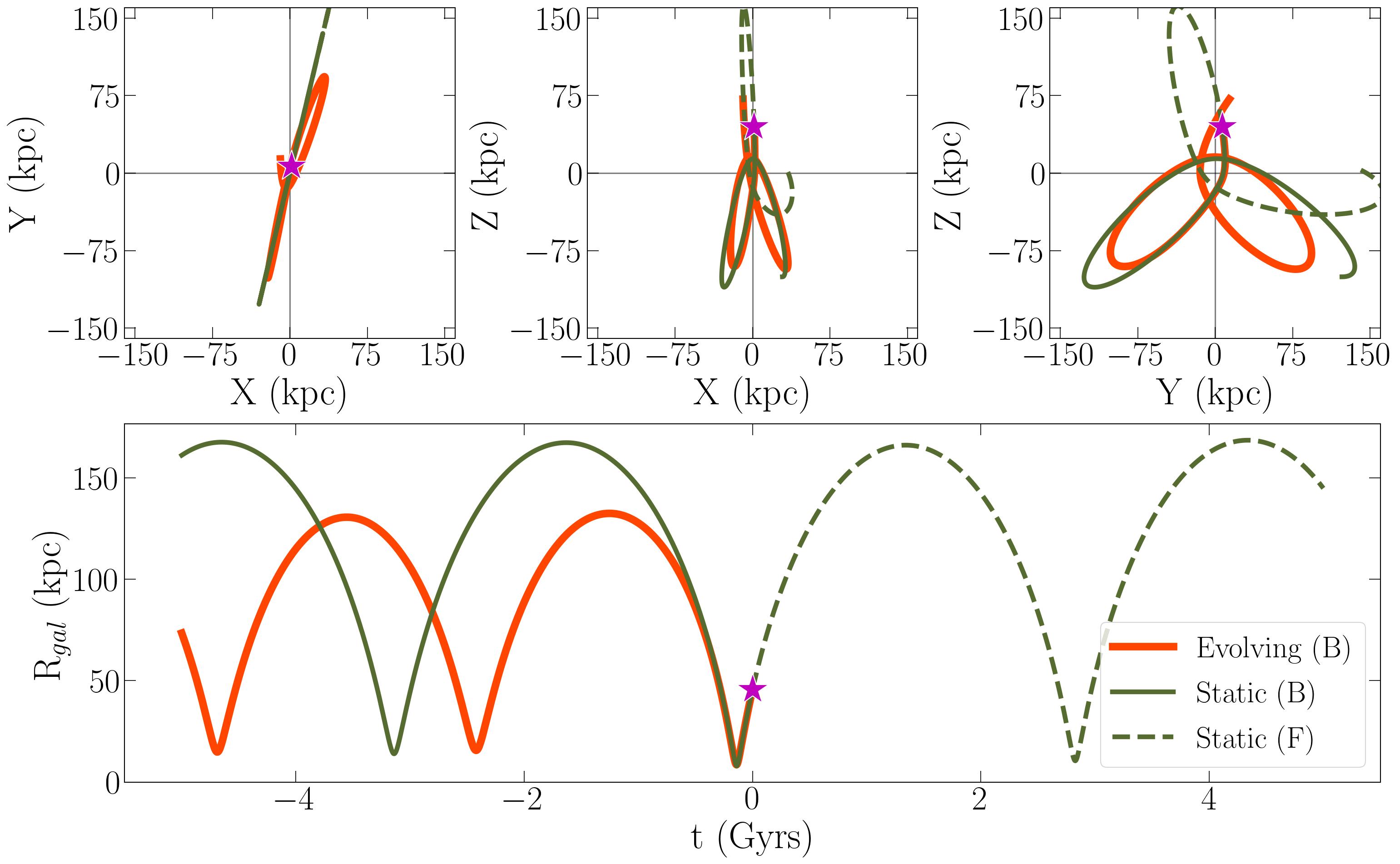}
    \caption{Boo3's orbit trajectories for both the static (olive) and time-evolving (orange) potentials. In both potentials, we integrate Boo3's orbit backwards by 5 Gyrs (shown as the solid lines). For the static potential only, we integrate the orbit forward by another 5 Gyrs (dashed line). The initial conditions of Boo3 are indicated as a magenta star.}
    \label{fig:orbit}
\end{figure}

The resulting orbits are shown in Figure \ref{fig:orbit} in Galactocentric coordinates. It is apparent in the bottom panel that Boo3 experiences more frequent pericentric passages in the evolving potential, and we note here that this detail may indeed prove relevant for future N-body simulations of Boo3’s disruption. We reserve a more detailed discussion of these implications for Section \ref{section:disc}.

% ’ equatorial coordinates and kinematics
We find that, for both potentials, the two orbits are practically identical within $\sim$20$^{\circ}$ of Boo3 such that either result can be used to examine the satellite's recent trajectory. In Sections~\ref{section:BHB_RGBs} and \ref{section:outer_MF}, we compare our observational results to the static potential orbit only for simplicity. A further discussion of the differences regarding the evolving potential will be presented later in Section \ref{section:disc} with our particle-spray analysis.

To explore the positions and velocity of Boo3 debris, we also approximate a ``particle-spray'' model (\citealt{fardal2015}) using the \texttt{AGAMA} framework. For this, we calculate the Galactocentric positions and velocities of the dwarf 3~Gyrs ago in its orbit, and use this timestep as the initial conditions for the progenitor in the model. At timesteps of 5 Myrs along the orbit, 2 massless particles are seeded at the satellite's Lagrange points and are assigned a typical escape velocity (derived from the self-gravity of Boo3, assuming a Plummer spherical potential; \citealt{plummer1911}). Thus the particles' velocities and positions account for both the orbit of the system at that timestep, in addition to the velocities imparted by the dwarf's own potential. These escaped particles are then integrated forward to the present day. The resulting distribution of particles produces an approximate stream model, allowing us to estimate the spatial and kinematic distribution of tidal debris without performing a full N-body simulation.

%%%%%%%%%%%%%%%%%%%%%%%%%%%%%%%%%%%%%%%%%%%%%%%%%%%%%%%%%%%%%%
\section{Data} 
\label{section:data}
This section describes the various data and catalogues used in this work. The first dataset used to study Boo3 and its putative stream is the list of stellar candidates identified in \textit{Gaia} eDR3 by \citet{jensen2024}. As part of this study, we also obtained the first narrow-band, CaHK imaging of Boo3 via the 3.6-meter Canada-France Hawaii Telescope (CFHT). These observations allow us to derive photometric metallicities across multiple fields, providing more stars with metallicity information than were obtained in previous spectroscopic studies. To conduct a wide-field search for individual stream members, we also use catalogues of stellar tracers (blue horizontal branch and red giant branch stars) that were identified using photometry from the Ultraviolet Near-Infrared Optical Northern Survey (UNIONS). Finally, we revisit the original matched filter evidence for stellar debris in the vicinity of Boo3 with additional photometric catalogues (SDSS and DELVE). Each of these datasets are discussed below.

\begin{figure*}
    \centering
    \includegraphics[width=0.9\textwidth]{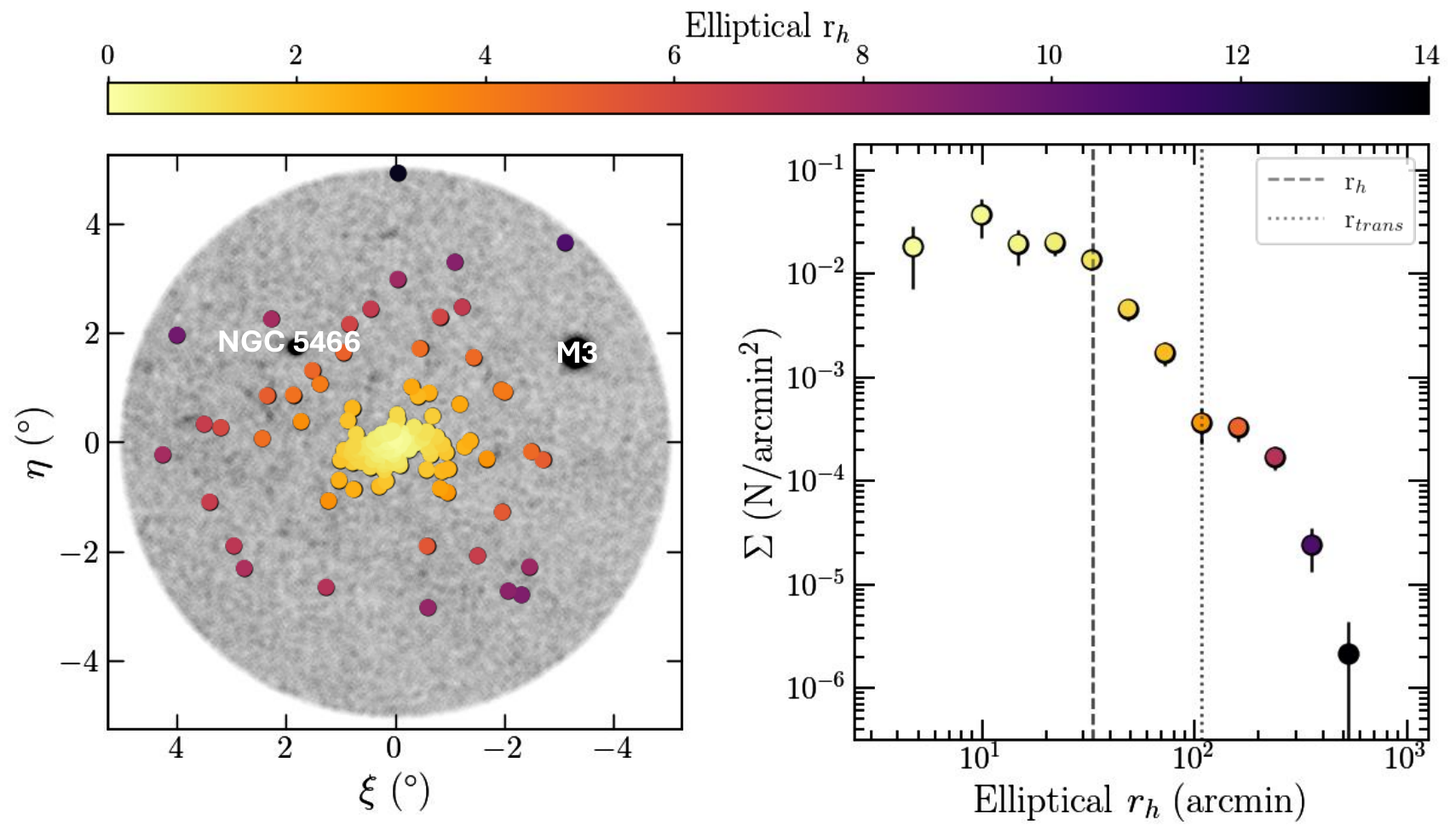}
    \caption{\textit{Gaia} candidate stars of Bo\"otes~III from \citet{jensen2024}. Data are colored according to their radial distance from the center, in units of elliptical half-light radii ($r_{h}$). \textbf{Left:} Tangent plane projection of candidates, overlain on MW field stars (grey). Globular clusters within the field are labeled in white text. \textbf{Right:} Stellar density profile of candidates as a function of elliptical $r_{h}$, in units of arcmin. For reference, we highlight $r_{h}$ and the transition radius ($r_{trans}$; from \citealt{jensen2024}) as the dashed and dotted vertical lines, respectively.}
    \label{fig:Boo3_Gaia}
\end{figure*}

\subsection{\textit{Gaia} Members from \citet{jensen2024}}
\label{section:gaiamems}

In \citet{jensen2024}, we applied a Bayesian-based algorithm to identify all likely stellar sources $-$ for \textit{all} MW dwarf spheroidals $-$ observed in \textit{Gaia} eDR3. The first dataset used in this work is the resulting candidate list for Boo3, obtained by \citet{jensen2024}. 
% \textit{Gaia}’s early Data Release 3 (eDR3; \citealt{gaia_edr3_2021})

To briefly summarize the methods of \citet{jensen2024}, the probabilities of membership for each star in a given field are calculated based on the star’s observed properties (spatial position including the presence of a low-density stellar substructure in the outskirts, color-magnitude, and proper motion) with respect to the dwarf and MW foreground models. We show in \citet{jensen2024} that, for stars with radial velocity measurements as an independent metric to confirm the purity of our samples, the algorithm performs exceptionally well at removing contamination by assigning appropriately low membership probabilities ($\lesssim$5\%) to MW foreground stars. Our dataset of stellar sources used in this present work are a subset of this catalogue, where the data are restricted to membership probabilities greater than 20\% (P$_{max}$ $\geq$ 20\%). In Figure~\ref{fig:Boo3_Gaia}, we showcase the physical extent of our Boo3 candidates in the tangent plane and stellar density profile (left and right panels, respectively). As indicated by the range of the color bar, these data span a significant radial distance (up to 14$r_{h}$) despite the algorithm not requiring radial velocity information to identify members.

\subsection{Follow-Up CaH\&K Photometry of Bo\"otes~III}
\label{section:pristine}

% Tidal interactions in dwarf galaxies are most clearly evident as morphological disturbances in the outskirts of their stellar distributions. However, these substructures are usually considerably lower in surface brightness than the main body of the progenitor (e.g., see the density profile of Pal-5 and its stream in \citealt{bonaca2020} and simulated profiles of the disruption of Sagittarius in \citealt{niederste-ostholt2012}) such that the signal of a stellar stream can be washed out by the dominating MW foreground. 

% While the previous study in \citet{jensen2024} alerted to an extended stellar population in Boo3, confirming each star's membership ideally requires spectroscopy to measure its radial velocity (RVs) and metallicity ([Fe/H]) and verify consistency with the progenitor. However, spectroscopic follow-up is observationally expensive as it requires substantially more observing time than photometric observations. Given the limited literature on Boo3 (e.g., \citealt{grillmair2009, carlin2009, correnti2009, carlin2018}), where even its half-light radius was only recently estimated (\citealt{moskowitz2020}) and its other structural parameters are not well-constrained, we pursued an alternative approach to study this system. Specifically, 

% As the \textit{Pristine} survey (\citealt{starkenburg2017}) does not presently have coverage near Boo3, we opted to obtain independent CaHK observations. 
% $\sim$1 square degree FOV
% to account for MegaCam's ``ears''

We obtained narrow-band Calcium H\&K (CaHK) photometric follow-up as a cost-effective way to estimate photometric metallicities across the field and distinguish Boo3 members from the more metal-rich MW foreground. These observations were conducted as part of a semester-long imaging program with CFHT, using the MegaCam imager (PI Errani; 22AF17). The completed observations total 6 central tiles covering $\lesssim$5$r_{h}$ of Boo3, where each field overlapped by 0.1$^{\circ}$. To achieve similar depths as \textit{Gaia} across as many fields as possible, each observation consisted of a single 912-second exposure per field. A total of 4.5 hours of observations were taken over the course of 4 nights (February 25, February 26, August 2, and August 3, 2022) in Semester 2022A. The raw data were pre-processed using CFHT's \texttt{elixir} pipeline (\citealt{magnier2004}) to conduct bias subtraction and apply flat-field corrections on the images. These data were then processed (astrometric calibration, extraction of photometric catalogue) following the steps described in \citet{martin2024}. 

We calibrate the CaHK photometry following the methodology of the \textit{Pristine} collaboration and the recommendations of \citet{martin2024}. In brief, two corrections are required before deriving photometric metallicities: a spatially varying correction across the MegaCam field-of-view (FOV), and an image-level zero-point correction accounting for varying observing conditions. The calibrated CaHK magnitude is therefore given by:

\begin{equation}
    {\rm CaHK}_{\rm calib} = {\rm CaHK}_{\rm uncalib} + zp(i) + FOV_{j}(x,y),
\end{equation}

\noindent where CaHK$_{\rm uncalib}$ is the raw instrumental magnitude, $zp(i)$ is the zero-point offset for image $i$, and $FOV_{j}(x,y)$ is the position-dependent FOV correction for observing run $j$.

We determine the spatial correction using \texttt{PhotCalib}$\footnote{\url{https://github.com/zyuan-astro/PhotCalib}}$, a specialized calibration algorithm developed by \citet{martin2024} for CaHK imaging at CFHT. \texttt{PhotCalib} uses a neural network trained on \textit{Pristine} data over multiple semesters, from 2015A to 2022B, to predict the CaHK magnitude correction as a function of pixel position in MegaCam's FOV. The training dataset includes star positions, raw CaHK magnitudes (CaHK$_{\rm uncalib}$), and \textit{Pristine-Gaia} synthetic CaHK magnitudes (CaHK$_{\rm syn}$)\footnote{\label{fn:pristinegaia}In the same manner as the predecessor catalogue (\citealt{starkenburg2017}), \textit{Pristine}'s newest data release relies on empirical relations between broad-band filters and narrow-band CaHK magnitudes to derive photometric metallicities. The survey's first data release (\citealt{martin2024}) includes a catalogue based on the CFHT observations of the \textit{Pristine} survey, as well as a new catalogue, dubbed the ``\textit{Pristine-Gaia} synthetic catalogue'', that is based on ``synthetic'' CaHK magnitudes, derived from \textit{Gaia} XP spectra. Both catalogues also now rely on \textit{Gaia} broad-band colors in lieu of SDSS. The XP-based ``\textit{Pristine-Gaia} synthetic'' catalogue is significantly shallower than the true \textit{Pristine} data due to the magnitude limit of XP observations in \textit{Gaia} DR3, but they provide a catalogue of all-sky CaHK magnitudes which we also use here to calibrate our ground-based CaHK observations.}. For our observations, we apply \texttt{PhotCalib} to our images (for either $j$ = 22Am02 or 22Am07 observing run) and obtain $FOV_{j}(x,y)$ for each source. Although the correction is derived using only stars cross-matched between our observations and the \textit{Pristine-Gaia} synthetic catalogue, we apply the resulting spatial correction to all stars in a given image including those fainter than $G~\lesssim~$21 mag.
% and synthetic CaHK magnitudes (CaHK$_{\rm syn}$) obtained from the \textit{Pristine-Gaia} synthetic photometric catalogue

% \footnote{\label{fn:pristinegaia}In the same manner as the predecessor catalogue (\citealt{starkenburg2017}), \textit{Pristine}'s newest data release relies on empirical relations between broad-band filters and narrow-band CaHK magnitudes to derive photometric metallicities. This new catalogue, dubbed the ``\textit{Pristine-Gaia} synthetic catalogue'', derives synthetic CaHK magnitudes from \textit{Gaia} XP spectra and utilizes \textit{Gaia} broad-band colors in lieu of SDSS. Though the data are shallower due to \textit{Gaia}'s magnitude limit, they provide a catalogue of all-sky CaHK magnitudes which we use here for calibrating our ground-based CaHK observations.}

We then compute the image-level zero-point offset, $zp(i)$, by determining the median difference between the spatially corrected CaHK magnitudes and their respective \textit{Pristine-Gaia}$\textsuperscript{\ref{fn:pristinegaia}}.$ CaHK$_{\rm syn}$ magnitudes. Extinction corrections for these data are implemented using \textit{Pristine}’s newly updated data processing pipeline (\citealt{martin2024}). This step is explained in more detail in Section \ref{section:inner}, where we also discuss how the photometric metallicities are calibrated.

\subsection{Tracer Stellar Populations Identified in UNIONS}
\label{section:tracers}

% We also search for Boo3’s putative stellar debris at large separations from the satellite by probing catalogues of stellar tracers. . These datasets consist of specific stellar populations 

We search for stellar debris at large separations from Boo3 by probing catalogues of specific stellar populations $-$ blue horizontal branch (BHB) and red giant branch (RGB) stars $-$ that have been identified with photometry from the Ultraviolet Near-Infrared Optical Northern Survey (UNIONS; \citealt{gwyn2025}). Both catalogues are products of UNIONS, a consortium that brings together several wide-field imaging programs of the northern sky. It combines the Canada-France Imaging Survey (CFIS) which is obtaining deep $u$- and $r$-band photometry from CFHT, deep $i$- and moderate-deep $z$-band imaging from Pan-STARRS, and deep $z$-band imaging through Wide Imaging with Subaru HyperSuprime-Cam of the Euclid Sky (WISHES) and $g$-band imaging through the Waterloo-Hawaii IfA $g$-band Survey (WHIGS). Although part of the original motivation was to supply optical imaging that complements the Euclid space mission, UNIONS operates as an independent collaboration whose primary goal is to maximize the scientific return of these large and deep surveys of the northern skies. 

\begin{figure}
    \centering
    \includegraphics[width=\linewidth]{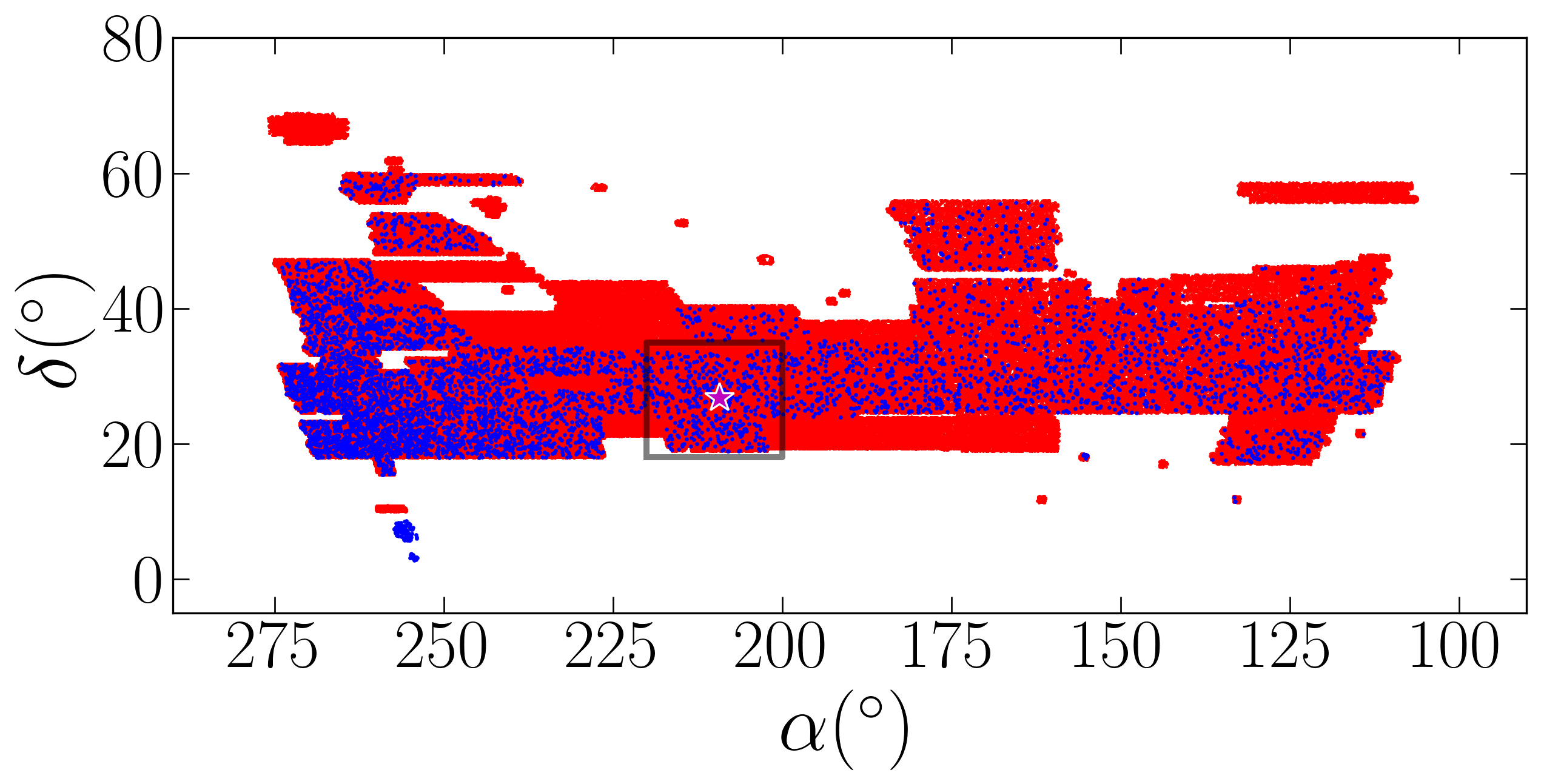}
    \caption{Equatorial coordinates of UNIONS BHBs (blue) and RGBs (red) stellar population catalogues for the largest contiguous area surrounding Boo3 (magenta star icon). The grey box highlights the region surrounding Boo3 selected for the sigma-clipping routine, described in Section \ref{section:outer}.}
    \label{fig:UNIONS}
\end{figure}

In these catalogues, every stellar source has well-constrained photometric distances, which (when paired with \textit{Gaia}) allow us to trace coherent kinematic signals across large Galactocentric distances. Figure~\ref{fig:UNIONS} presents the subsample of BHBs (blue) and RGBs (red) used in this study, where the footprint is motivated by selecting the largest continuous region containing Boo3. These catalogues originate from early observations contributing to UNIONS, specifically the $u$-band from CFIS (\citealt{ibata2017}). We note that there are indeed differences in coverage between the two catalogues, due to the on-going observations between their publication in 2018 and 2019. Jointly, however, the two datasets provide sufficient coverage of Boo3 (highlighted in the figure with a magenta star), and have already proven useful to trace MW substructure (e.g., \citealt{jensen2021}).

The first stellar tracer catalogue derived from UNIONS/CFIS is the BHB catalogue of \citet[hereafter \citetalias{thomas2018}]{thomas2018}. BHB stars are useful halo tracers because of their relatively stable absolute magnitudes (M$_g \sim$ 0.5 $-$ 0.7 mag; \citealt{deason2011}) from which heliocentric distances can be estimated to $\sim$10\% precision. Their intrinsic brightness allows them to be detected at large distances ($>$100 kpc), making them valuable tracers of MW substructure. \citetalias{thomas2018} used CFIS $u$- and Pan-STARRS 3$\pi$ (PS1~3$\pi$) $griz$-bands to (i) photometrically identify A-type stars in the CFIS footprint and (ii) classify them as either BHBs or blue stragglers using machine learning methods calibrated against spectroscopic samples. The final catalogue contains $\sim$10,200 BHB candidates, with an estimated contamination rate below $\sim$25\% and heliocentric distances extending to $\sim$220 kpc.

Because BHB stars are intrinsically rare, we supplement this sample with the UNIONS ``Dwarfs/Giants'' catalogue of \citet[hereafter \citetalias{thomas2019}]{thomas2019}. This catalogue uses the combined CFIS-PS1~3$\pi$-\textit{Gaia} photometry to classify stars as either ``dwarf'' or ``giant'' and provides estimates of photometric metallicity and absolute magnitude for $\sim$12.8 million sources. We use the giant candidates as a complementary, higher-density tracer of halo substructure. Following the recommendations of \citetalias{thomas2019}, we retain stars with $P_{\rm giant} \geq 0.5$ (136,000 stars) and apply recommended quality cuts. In particular, we remove extended sources by requiring $|r_{\rm PSF} - r_{\rm ap}| < 0.05$ mag, and restrict the sample to stars with well-constrained photometric errors by requiring $\delta M_{G,{\rm pred}} \lesssim 0.5$ mag, where $\delta M_{G,{\rm pred}}$ includes photometric and systematic uncertainties added in quadrature.

The original datasets in \citetalias{thomas2018} and \citetalias{thomas2019} utilized the second data release from \textit{Gaia} (\citealt{gaia_collab_2018}), and therefore already contain proper motion and parallax information. Though the photometric measurements between data releases does not vary significantly, both a star's parallax and proper motion are greatly enhanced with additional observation epochs such that newer data releases will continue to increase in \textit{Gaia}'s accuracy. For this reason, we cross-match both the BHBs and RGBs samples to \textit{Gaia}’s third data release (DR3; \citealt{gaia_collab_dr3_2023}). To ensure we avoid nearby disk stars (\citealt{lindegren2018}) and simultaneously account for \textit{Gaia}’s systematic parallax offset ($-$0.017 mas; \citealt{lindegren2021}), we apply the following cut to both the BHB and RGB datasets:

% \begin{equation}
%     \frac{1}{\pi + 0.03 \; \text{mas}} > 5 \; \text{kpc}
% \end{equation} 

\begin{equation}
    \frac{1}{\pi + 0.017 \; \text{mas}} > 5 \; \text{kpc}
\end{equation} 

\noindent This restriction ensures we remove all stars with distances of $<$ 5 kpc, or within the Solar neighborhood.

\subsection{Matched Filter Searches Using SDSS and DELVE Photometric Catalogues}
\label{section:surveys}

% and compare to our observations in this work

% Using the MF method (explained in more detail in Section \ref{section:outer_MF}) achieves two goals: we can (i) attempt to replicate results of the initial detection, and (ii) make our own characterizations of the dwarf and/or its stream. Furthermore, by applying this analysis to multiple photometric catalogues, we should be able to better inspect observational artifacts that may have resulted in spurious detections. 

To conduct a wide-field search for stellar debris of Boo3, we next apply the matched filter (MF) technique on two additional photometric catalogues. For both, we queried the respective databases for stars located in a substantial area surrounding Boo3 corresponding to $154^{\circ} \leq \text{RA} \leq 230^{\circ}$ and $+7^{\circ} \leq \text{Dec} \leq +43^{\circ}$. This region encompasses most of the area included in the original detection. We do not find it necessary to extend this area given that the field becomes overwhelmed by the Sagittarius stream in the West (RA below $\sim$190$^{\circ}$). We rely on both in order independently confirm any detections in the MF map, and to assess artificial/spurious features in the MF density maps. 

The first catalogue used is photometry from the Sloan Digital Sky Survey (SDSS; \citealt{york2000}). SDSS is a large-scale astronomical survey that achieves dual-hemisphere coverage via two 2.5-meter telescopes $-$ the Sloan Foundation telescope at Apache Point Observatory in New Mexico, and the du Pont telescope at Las Campanas Observatory in Chile. Both Boo3 and Styx were originally discovered in SDSS by \citet{grillmair2009}; here, we revisit the detection by conducting our own MF analysis on a later data release. For this work, we acquired photometric data covering a broad area from the most recent SDSS data release, DR17 (\citealt{APOGEE_DR17_abdurrouf2022}), using the \texttt{CasJobs}\footnote{\url{https://skyserver.sdss.org/CasJobs/}} SQL interface. Specifically, we downloaded photometric measurements in the $g$-, $r$-, and $i$-bands and their associated errors over a wide field of northern sky. This data is downloaded from SDSS DR17’s \texttt{star} catalogue, which comprises high-confidence stellar sources identified internally by SDSS. To ensure good quality photometry, we removed stars that are flagged with poor measurements (magnitude or magnitude error = $-$9999.0). We further limit the catalogue by recalculating the 5-$\sigma$ depths in each band. For each filter, we compute the median photometric error in each band as a function of magnitude, and interpolate this function to determine the depth corresponding to a median photometric error of 0.2 mag. Only sources brighter than this limit are retained; in practice, our SDSS 5-$\sigma$ depths in $gri$-bands are 23.2, 22.8, and 22.3 mag, respectively. 
% We limit the catalogue further by determining a photometric depth in each filter, and retaining only data whose magnitudes are brighter than these limits. To do so, we determine the median photometric error in each band as a function of magnitude, and interpolate this function for where the median photometric error is equal to a threshold of 0.2 mag. In practice, this means that SDSS' 5-$\sigma$ depths in $gri$-bands are 23.2, 22.8, and 22.3 mag, respectively. 

The second photometric catalogue used here is from the DECam Local Volume (DELVE; \citealt{drlica-wagner2022}) survey, which is deeper than SDSS\footnote{Unfortunately, we are unable to use the UNIONS catalogue here, as Boo3 lies outside the main multi-band survey footprint and only has $u$-band coverage.}. The DELVE survey is another optical photometric catalogue focused on obtaining \textit{griz}-band photometry for sources in the southern hemisphere (declinations below $<$~30$^{\circ}$, excluding the Galactic disk). Taken using the 4-meter Blanco Telescope at Cerro Tololo Interamerican Observatory in Chile, this imaging survey aims to uncover newly discovered faint MW satellites and characterize their properties (e.g., \citealt{mau2020, martinez-vazquez2021, cerny2021_delve2, cerny2021_eri4}). For our analysis, we download a dataset of $g$- and $r$-band sources from the second DELVE data release (DR2; \citealt{drlica-wagner2022}). While it would be beneficial for the MF method to also use DELVE's $i$-band, these observations are not continuous this far north; as such, we only utilized $g$- and $r$-band magnitudes from DELVE. To ensure the dataset consists only of the most reliable stellar sources, we follow several criteria recommended in \citet{drlica-wagner2022} for DELVE DR2:

\begin{itemize}
    \item The total sample is first limited to sources where \texttt{extended\_class}~$\leq$~1. This flag indicates the confidence of a source to be a star ($\leq$1) or galaxy ($>$1), as identified using \texttt{SourceExtractor}.
    \item We remove stars with poor photometric measurements by limiting our sample to stars with \texttt{mag\_PSF} $<$ 99 and \texttt{magerr\_PSF} $<$ 99. This flag indicates sources with poor photometry (and uncharacterized photometric errors) according to the DELVE collaboration. 
    \item We limit the catalogue by our own estimates for the photometric depths in each band. Given the 5-$\sigma$ depths we calculate, DELVE's $g$- and $r$-bands are limited to 24.0 and 23.3 mag, respectively.
\end{itemize}

Finally, we correct both SDSS and DELVE photometric catalogues for extinction using the extinction coefficients from \citet{schlafly_finkbeiner2011} and the \texttt{python}-wrapped package \texttt{dustmaps} (\citealt{green_dustmaps2024}).

% Our estimates for DELVE's median 5-$\sigma$ depth is slightly deeper than SDSS: in $gri$, the limits are 24.3, 23.9, and 23.5 mag, respectively. Compared to the shallower limits of SDSS (23.2, 22.8, and 22.3 mag), DELVE provides us with an independent dataset to confirm any detections, as well as to assess artificial/spurious features in the MF density maps. 

% Below, we discuss the catalogues used in the MF analysis and what recommended calibrations were applied to these datasets. 

% \href{https://skyserver.sdss.org/CasJobs/}{\texttt{CasJobs}}

%%%%%%%%%%%%%%%%%%%%%%%%%%%%%%%%%%%%%%%%%%%%%%%%%%%%%%%%%%%%%%
\section{Analysis of the Main Body of Bo\"otes~III}
\label{section:inner}

\subsection{Calibrating Photometric Metallicities}

To calibrate our CaHK observations and obtain photometric metallicities, we rely on the methods of the ``Pristine dwarf galaxy'' program (e.g., \citealt{longeard2018, longeard2020, longeard2022, longeard2023}), a subset of the larger \textit{Pristine} collaboration, and implement the updated methodology described in \citet{martin2024}. Deriving photometric metallicities for stars in and around Boo3 allows us to: (i) verify Boo3's previous metallicity estimate, and (ii) assess each star's membership to the dwarf, especially given that Boo3 is more metal-poor (${\rm [Fe/H]} \sim -2.1$ dex; \citealt{carlin2009}) than the MW disk (${\rm [Fe/H]} \sim -0.5$ dex, albeit with considerable spread; \citealt{helmi2020}). Briefly, the algorithm derives photometric metallicities from empirical relations between narrow-band CaHK magnitudes, broad-band colors, and intrinsic stellar properties such as temperature and surface gravity. These relations are mapped in color-color space using a grid of synthetic spectra, allowing photometric metallicities to be estimated from the observed photometry of individual stars. In this work, we use the newest iteration of this method, which is optimized for \textit{Gaia} broad-band magnitudes (\citealt{martin2024}). Because  \textit{Gaia}'s passbands are very broad, determining an extinction correction is complex: the extinction in a given filter ($A_f$) depends not only on a star's position on the sky, but also on its underlying stellar properties, including temperature, $\log(g)$, and metallicity. To account for this dependence, \citet{martin2024} developed an algorithm that iteratively solves for the extinction and stellar properties simultaneously. The method first establishes the dependence of the extinction coefficients ($k_f$) on effective temperature (T$_{eff}$) and metallicity ([Fe/H]) in the \textit{Gaia} bands, where $A_f = R_f \times E(B-V)$ and $R_f$ is a function of $k_f$. For an individual star, the algorithm first calibrates $A_f$ provided with (i) the star's reddening ($E(B-V)$) from the \citet{schlegel1998} dust maps, and (ii) an initial estimate of T$_{eff}$ from the observed $(B_P - R_P)$ color. The first guess for the photometric metallicity is then calculated, and the process is repeated with the updated [Fe/H] estimate. This process is repeated iteratively until the values converge.

The final parameters provided from the algorithm are then $k_{f}$, T$_{eff}$, and [Fe/H]$_{phot}$ for each star. Uncertainties for the photometric metallicity are determined through a Monte Carlo (MC) simulation, where 100 samples are drawn from each star’s photometric uncertainties. The final photometric metallicity is then reported as the median of this estimate, with errors corresponding to the 16$^{th}$ and 84$^{th}$ percentiles of the probability density function. 

To calibrate the photometric metallicities for our Boo3 observations using the updated \textit{Pristine} pipeline from \citet{martin2024}, we first cross-match our CaHK photometry with broad-band magnitudes from \textit{Gaia} DR3. The resulting cross-matched dataset is identical to the sample from \citet{jensen2024}, meaning the data also now include membership probabilities. Initially, our dataset includes CaHK-derived metallicities for 13,258 stars across all 6 observed fields.

To retain only the most reliable of these data, we apply several quality cuts recommended by \citet{martin2024}. We remove stars with metallicities at the edges of the synthetic grid, retaining only those within $-4.0 < \text{[Fe/H]} \leq 0.0$. Additionally, we limit the sample to stars whose MC sampled photometry largely remain within the color-color space of the metallicity model (\texttt{mcfrac\_Pristine} $\geq$ 0.8). Following \textit{Gaia} quality criteria, we further restrict our sample to sources with reliable astrometric and photometric measurements, specifically requiring \texttt{ruwe} $<$ 1.4 (or flagged by \texttt{F\_BEST}~=~1 as done in \citealt{jensen2024}). These criteria yield a final sample of 2040 stars.

\subsection{Comparison of \citet{jensen2024} Boo3 Members}

\begin{figure*}
    \centering
    \includegraphics[width=1\textwidth]{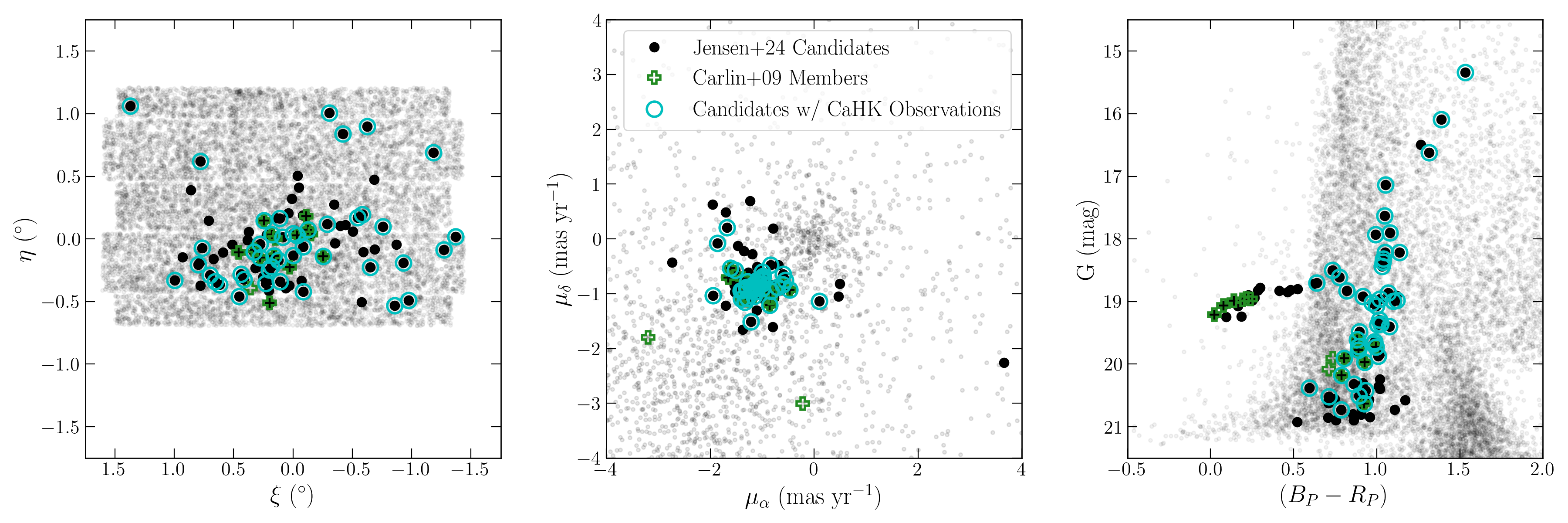}
    \caption{Tangent position centered on Bo\"otes~III (left), proper motions (center), and CMD (right) of stars from our observing runs. Black points in this figure indicate the \textit{Gaia} candidate members (as in the color-coded points in the left panel of Figure \ref{fig:Boo3_Gaia}), but we only plot those within the same region of sky as the targeted fields. Cyan circles highlight the \textit{Gaia} candidate members with new CaHK observations. For comparison, we also plot the \citet{carlin2009} spectroscopic members as green plus (\textbf{+}) icons. The remaining foreground data from \textit{Gaia} are shown in grey.}
    \label{fig:Pr_Gaia_Mems}
\end{figure*}

Given that the previous estimate of Boo3’s metallicity dispersion is particularly wide for a dwarf galaxy ($\sigma_{\rm Fe/H}$ = 0.6; \citealt{carlin2009}), we find that selecting a sample of Boo3 stars based solely on a range of photometric metallicities is insufficient to effectively discriminate dwarf galaxy membership. Out of the current sample, there are 661 stars whose metallicities are contained within 1-$\sigma$ of Boo3’s range ($-$2.7 to $-$1.5 dex). Therefore, we incorporate an additional membership criterion (P$_{max} \geq$ 0.2) from \citet{jensen2024}, narrowing our sample to 39 stars consistent with membership to Boo3. The final sample of Boo3 stars with CaHK-derived photometric metallicities are indicated with cyan open circles in Figure \ref{fig:Pr_Gaia_Mems}. For comparison, all cross-matched \textit{Gaia} stars in the 6 observed fields appear in grey, while stars specifically from \citet{jensen2024} with consistent membership probabilities are indicated in black. We also plot stars that have been spectroscopically-identified as members in \citet{carlin2009} (cross-matched to \textit{Gaia}) as green plus (\textbf{+}) icons.

% This first difference in samples is solely due to the \textit{Pristine} CaHK magnitude calibration, which only derives photometric metallicities for stars with 0.5 $<$ ($B_{P}$ $-$ $R_{P}$)$_{0}$ $<$ 1.5 due to the ionization of calcium in bluer, hotter stars resulting in intrinsically weaker CaH\&K absorption lines.
The differences between the \citet{jensen2024} Boo3 data (black) and our photometric metallicity sample (cyan) in Figure \ref{fig:Pr_Gaia_Mems} arise due to three factors. As shown clearly in the right panel, our photometric metallicity sample is strictly limited to the RGB, and is particularly sparsely populated at the faintest magnitudes. The first discrepancy between samples arises because the \textit{Pristine} CaHK magnitude calibration only derives photometric metallicities for stars with 0.5 $<$ ($B_{P}$ $-$ $R_{P}$)$_{0}$ $<$ 1.5; bluer, hotter stars have weaker CaH\&K absorption lines due to calcium ionization, and as such they are excluded from the analysis. This limitation therefore excludes the entirety of the horizontal branch from the CaHK photometric metallicity sample. We also find that the sparse sampling of stars at $G~>~$20 mag is caused by the recommended limit on \texttt{mcfrac\_Pristine}; these stars are excluded due to their MC sampled photometric errors which fall beyond the color-color space grid. The last difference between samples is the presence of a bright RGB star in \citet{jensen2024} but not in our CaHK data, indicated as the only RGB (in black) with no surrounding cyan indicator, found in the CMD panel at at ($B_P - R_P$,~$G$) =  (1.3 mag, 16.5 mag). We find that this star was excluded from the calibration pipeline, due to its location precisely on a CCD chip gap in the MegaCam imager during our observations.
% (1.75, 16.6) (\textcolor{red}{CHECK THIS WITH THE DATA})

Previous spectroscopic observations of Boo3 by \citet{carlin2009} yielded metallicities and radial velocities for 193 stars in the vicinity of the dwarf, with 20 stars flagged as members based on similar radial velocities. After cross-matching with \textit{Gaia}, 16 of these remain and are marked as green \textbf{+} icons in Figure \ref{fig:Pr_Gaia_Mems}. However, 2 of these stars have membership probabilities inconsistent with Boo3 due to discrepant proper motions (visible as outliers in the central proper motion panel), and an additional 7 HB stars are excluded due to the aforementioned color limitations for the \textit{Pristine} algorithm. This leaves only 7 spectroscopic RGB members for direct comparison to our \textit{Pristine} photometric metallicities.

\subsection{Bo\"otes~3's Metallicity Distribution Function}
\label{section:MDF}

Given Boo3's broad metallicity range derived in previous works and lack of recent spectroscopic follow-up, we independently estimate the dwarf’s mean metallicity ([Fe/H], or $\langle x_{Fe/H} \rangle$) and metallicity dispersion ($\sigma_{\rm Fe/H}$) from our larger photometric metallicity sample. Adopting an approach similar to \citet{walker2006}, we use Maximum Likelihood Estimation in a Markov Chain Monte Carlo (\citealt{hastings1970}; MCMC) framework to derive the best-fit parameters provided our observations and their associated errors. The log-likelihood function to be maximized is defined as:

\begin{equation}
    \begin{split}
        \ln(\mathcal{L}) = -\frac{1}{2} \sum_{i = 1}^{N} \ln(\sigma_{i}^{2} + \sigma_{Fe/H}^{2}) \\ -~\frac{1}{2} \sum_{i = 1}^{N}\frac{(x_{i} - \langle x_{Fe/H} \rangle)^{2}}{\sigma_{i}^{2} + \sigma_{Fe/H}^{2}} - \frac{N}{2} \ln(2 \pi)
    \end{split}
\end{equation}

% \begin{equation}
%     % \ln(\mathcal{L}) = -\frac{1}{2} \sum_{i = 1}^{N} \ln(\sigma_{i}^{2} + \sigma_{Fe/H}^{2}) - \frac{1}{2} \sum_{i = 1}^{N}\frac{(\bar{x}_{i} - \bar{x}_{Fe/H})^{2}}{\sigma_{i}^{2} + \sigma_{Fe/H}^{2}} - \frac{N}{2} \ln(2 \pi)
%     \ln(\mathcal{L}) = -\frac{1}{2} \sum_{i = 1}^{N} \ln(\sigma_{i}^{2} + \sigma_{Fe/H}^{2}) - \\ \frac{1}{2} \sum_{i = 1}^{N}\frac{(x_{i} - \langle x_{Fe/H} \rangle)^{2}}{\sigma_{i}^{2} + \sigma_{Fe/H}^{2}} - \frac{N}{2} \ln(2 \pi)
% \end{equation}

\noindent where $x_{i}$ and $\sigma_{i}$ represent each star’s observed photometric metallicity and uncertainty, respectively. We evaluate the log-likelihood function above using the \texttt{emcee} python package (\citealt{foreman-mackey2013}) and solve for $\langle x_{\text{Fe/H}} \rangle$ and $\sigma_{\rm Fe/H}$. The only priors placed on the posterior distributions are that the metallicity must range only between $-$4.0 $<$ $\langle x_{\text{Fe/H}} \rangle$ $\leq$ 0.0 and the dispersion must be greater than 0~dex. For the mean metallicity and dispersion of the system, we estimate $\langle {\rm{Fe/H}} \rangle$~=~$-$2.1~$\pm$~0.1 and $\sigma_{\rm Fe/H}$~=~0.2~$\pm$~0.1, respectively. 
% We then solve for the combination of $\langle x_{\text{Fe/H}} \rangle$ and $\sigma_{[\text{Fe/H}]}$ that maximizes the log-likelihood function above, using the \texttt{emcee} python package (\citealt{foreman-mackey2013})

% The resulting posterior distribution functions (PDFs) are presented in the top panel of Figure \ref{fig:mdf}. 

\begin{figure*}
    \centering
    \includegraphics[width=\textwidth]{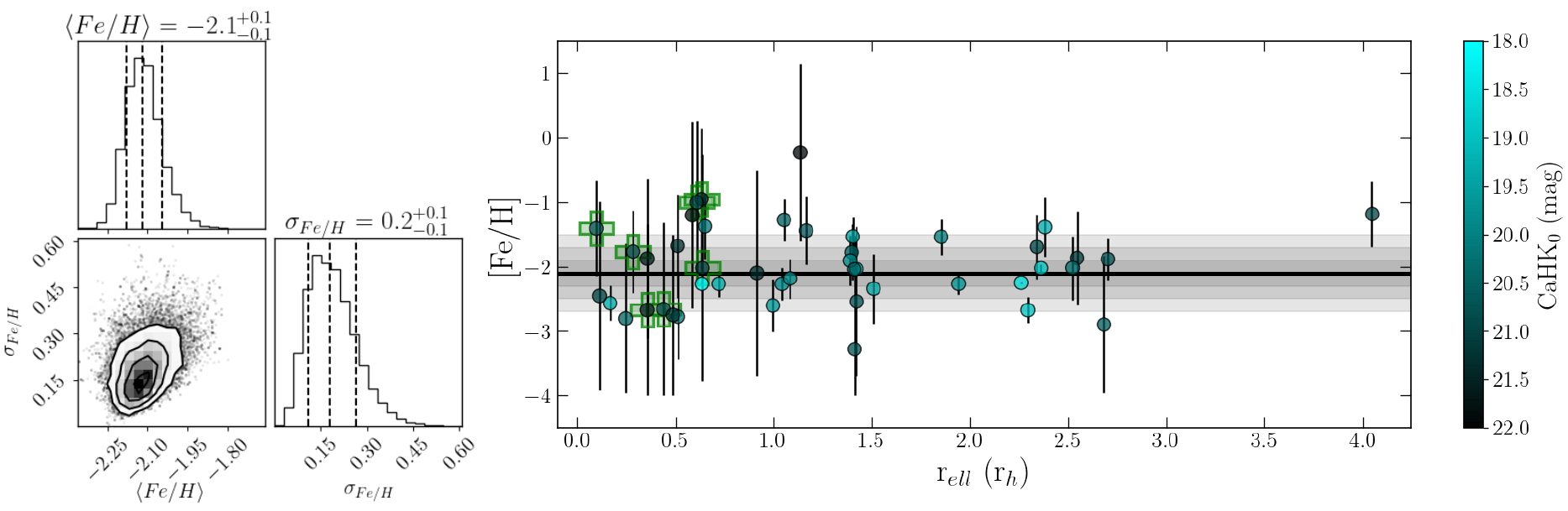}
    \caption{Metallicity distribution function of our Boo3 observations. The left panel presents the results of the MCMC for the metallicity and metallicity dispersion. The right shows photometric metallicities and errors for stars we consider members of Boo3 (probabilities $>$ 20\%) as a function of elliptical half-light radius ($r_{ell}$) in units of $r_{h}$. Each star is colored corresponding to its CaHK magnitude, and previous spectroscopic members of Boo3 from \citet{carlin2009} are indicated as green \textbf{+} icons. To showcase the spread in metallicity of these observations compared to our MCMC estimates for the dwarf, we also show Boo3's mean metallicity ([Fe/H]) estimate as the black horizontal line, as well as the ranges indicated by 1$\times$, 2$\times$, and 3$\times$ our estimate for the metallicity dispersion ($\sigma_{\rm Fe/H}$) shaded in the grey bands. Note that our estimate for the dispersion is only $\sigma_{\rm Fe/H}$ = 0.2 dex relying on the photometric metallicities of 39 total stars, while the previously reported estimate of 0.6 dex relied on only 20 (\citealt{carlin2009}).}
    \label{fig:mdf}
\end{figure*}

From this analysis, we find agreement between the mean metallicity derived from our CaHK-derived sample and previous spectroscopic measurements. However, the dispersion we derive is significantly smaller $-$ three times less than reported by \citet{carlin2009}. In direct comparison with the subset of 7 spectroscopically confirmed RGBs from \citet{carlin2009}, we find broad consistency between metallicity estimates within 1-$\sigma$, though we note our errors are substantially larger. The inflated uncertainties for these stars arise due to their faint magnitudes ($G >$~19.5 mag) and inherently larger photometric errors, which propagates to our CaHK-derived metallicities. Nevertheless, we observe that much of our sample has much smaller errors overall and furthermore are robust due to their membership probabilities from \citet{jensen2024}.

Figure \ref{fig:mdf} illustrates our findings from the MCMC analysis, where the left panel displays the MCMC posterior distributions for [Fe/H] and $\sigma_{\rm Fe/H}$. The right panel we present out CaHK-derived photometric metallicities as a function of elliptical half-light radius, color-coded by CaHK magnitudes. The black horizontal line indicates the our new estimate for the mean metallicity, with grey shaded regions indicating 1$\times$, 2$\times$, and 3$\times$ the new dispersion. Our dataset significantly expands the available metallicity information (shown as green \textbf{+} icons), reaching distances up to 4$r_h$. We therefore consider our estimate for the metallicity and dispersion to be robust and reliable for subsequent analyses. In the next section, we leverage this updated information to investigate the stellar tracer catalogues for Boo3 debris, where the updated metallicity dispersion will provide better constraints on our selection of RGB candidates.

%%%%%%%%%%%%%%%%%%%%%%%%%%%%%%%%%%%%%%%%%%%%%%%%%%%%%%%%%%%%%%
\section{Pushing Out to Large Distances: Searching for Stellar Debris of Bo\"otes~III}
% \section{Outer Study of Bo\"otes 3: a Search for Styx}
\label{section:outer}

The goal of this section is to search for stellar debris associated with Boo3. In Section~\ref{section:BHB_RGBs}, we first probe the BHB and RGB datasets derived from UNIONS to obtain putative stream candidates. Then in Section~\ref{section:outer_MF}, we explore a significant area of sky for evidence of a stream-like structure by creating matched filter maps from SDSS and DELVE and compare previous results to our own.

\subsection{Kinematic Detection Method for Stream Identification}
\label{section:BHB_RGBs}

% We search for extended debris of Boo3 using the BHB and RGB stellar tracer catalogues. To identify putative substructure of the dwarf, we first 
Using the UNIONS stellar tracer catalogues, we now search for extended debris of the Boo3 dwarf. We first take a subsample of giants whose properties are broadly consistent to that of the progenitor, and then probe this data for kinematic signatures. Trends in proper motions (and distances) across the sky are a useful indicator of a shared dynamical origin and may uncover hidden substructures such as a coherent stellar stream. For this analysis, we use the same sigma-clipping routine from \citet{jensen2021} to identify stellar stream candidates. The benefit of this method over selecting stars by eye is that we are able to make an independent detection that does not rely on prior knowledge, such as Boo3’s orbit or the reported location of Styx in \citet{grillmair2009}, which may unintentionally bias our detection.

\begin{figure}
    \centering
    \includegraphics[width=\linewidth]{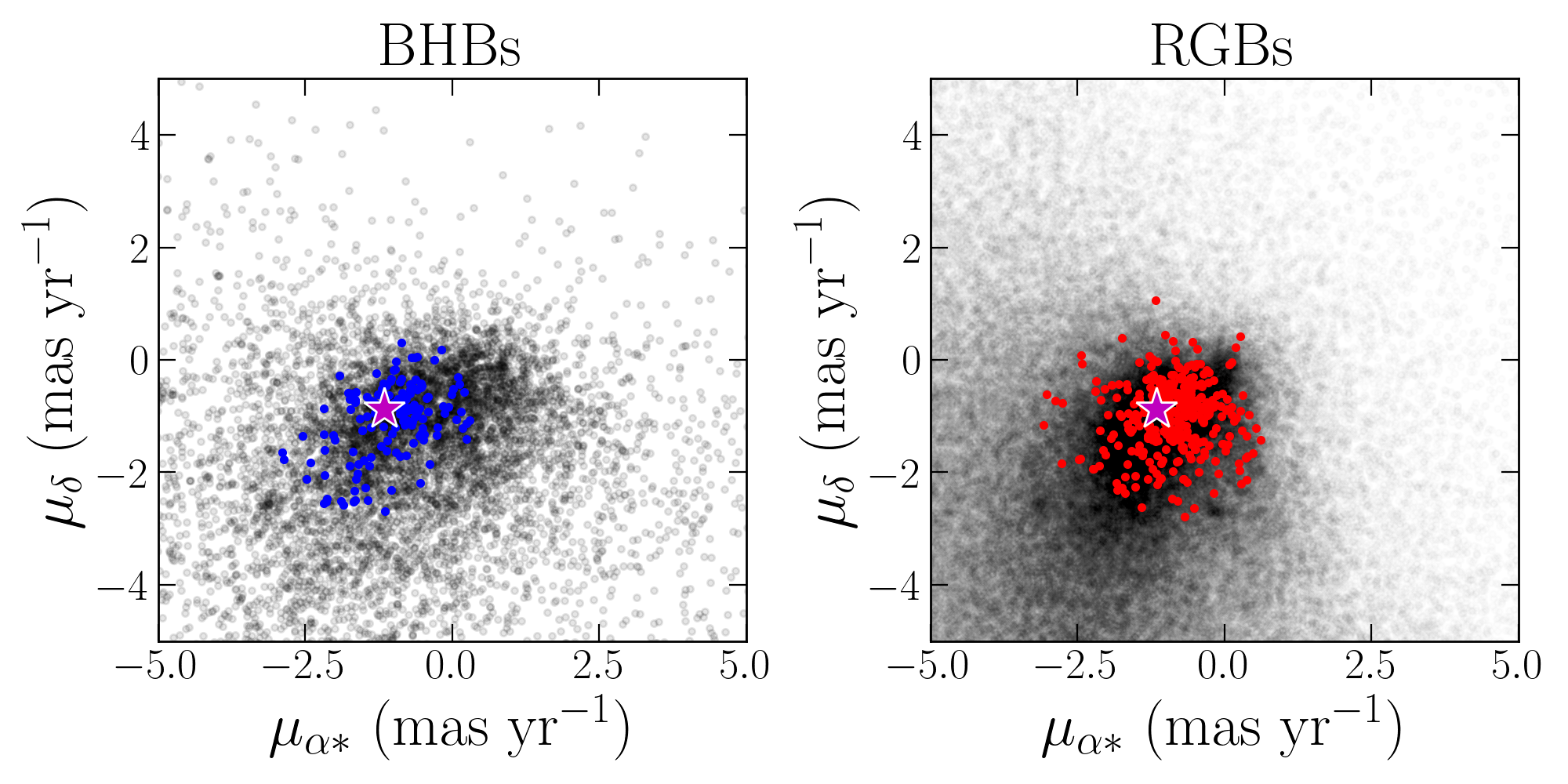}
    \caption{Proper motion space of the \citetalias{thomas2018} (left panel) and \citetalias{thomas2019} (right panel) catalogues represented as grey points. The data from these catalogues that are retained for the sigma-clipping routine are indicated by the colored points; blue for the BHB sample and red for the RGBs. The systemic motion of Boo3 is indicated as the magenta star icon.}
    \label{fig:pmspace_sample}
\end{figure}

To first isolate stars consistent with being members of Boo3, we apply the following restrictions uniformly to the BHB and RGB catalogues from UNIONS. We first limit the data spatially to a region of sky between $200^{\circ}\leq\text{RA}\leq220^{\circ}$ and $+18^{\circ}\leq\text{Dec}\leq+35^{\circ}$, indicated with a grey outline in Figure \ref{fig:UNIONS}. This boxed area mainly serves to ensure the coverage between the catalogues is approximately continuous, while also encompassing a large enough area around Boo3 to detect any lengthy substructure. We then exclude data from this set if their proper motions significantly vary from Boo3, by only keeping stars whose proper motions are within a radius of 2~mas~yr$^{-1}$ to that of the dwarf. This measure allows us to remove many MW disk stars that are inconsistent with membership based on their kinematics, as shown in Figure \ref{fig:pmspace_sample}. The last shared restriction is that we limit the data to only stars with heliocentric (photometric) distances between 35 and 80~kpc. This range allows us to retain debris at or near the distance of Boo3 ($\sim$46~kpc), while also permitting us to explore potential distance gradients in the putative stream.

The RGB catalogue includes an additional parameter than that of the BHBs, which are photometric metallicities. In addition to the aforementioned measures, we further limit the RGB dataset to a range of metallicities consistent to Boo3 using our updated metallicity dispersion from Section \ref{section:MDF}. The final sample is restricted to $-2.3 < \text{[Fe/H]}_{phot} < -1.9$~dex, corresponding to the mean metallicity of Boo3 $\pm 1$-$\sigma_{\rm Fe/H}$.

The next stage of this analysis is to apply the sigma-clipping routine from \citet{jensen2021} to isolate a subset of stars whose kinematics and positions are consistent with Boo3 debris. If a coherent structure is present in the data, this procedure should yield a sample of extended Boo3 debris indicative of stars forming a stellar stream. The steps of the routine are listed below, as follows:

\begin{enumerate}[label=(\roman*)] 
    \item The first step of the algorithm is to plot the proper motion in RA ($\mu_{\alpha*}$) versus RA ($\alpha$) for the dataset. We take a simple linear fit to the data, where the weights of this fit are the errors in each star’s proper motion in RA;
    \item Next, we identify all stars in the subsample whose $\mu_{\alpha*}$ measurements are 3-$\sigma$ within this fitted line. We then generate a new linear fit to this smaller subsample;
    \item We take the one-degree polynomial fit from the previous step and compare it to the total dataset of stars we initially put into the algorithm. We then retain a new subsample of stars whose $\mu_{\alpha*}$ are consistent with this new fit (within 3-$\sigma$); 
    \item We repeat step (iii) step until convergence, such that the generated fit and sample of stars retained no longer vary;
    \item For the subsample of stars that survive the sigma-clipping in $\mu_{\alpha*}$ vs $\alpha$, we repeat the process with the proper motion in Declination ($\mu_{\delta}$) vs $\alpha$\footnote{Previously in \citet{jensen2021}, we conducted this second sigma-clipping routine in $\mu_{\delta}$ vs Declination ($\delta$). However, Boo3's orbit in $\mu_{\delta}$ vs $\delta$ is more complex than can be described with a simple linear fit. For this reason, we opted to change the x-axis to $\alpha$ for the second round of sigma-clipping.}.
    
\end{enumerate}

\begin{figure*}
    \centering
    \includegraphics[width=0.8\textwidth]{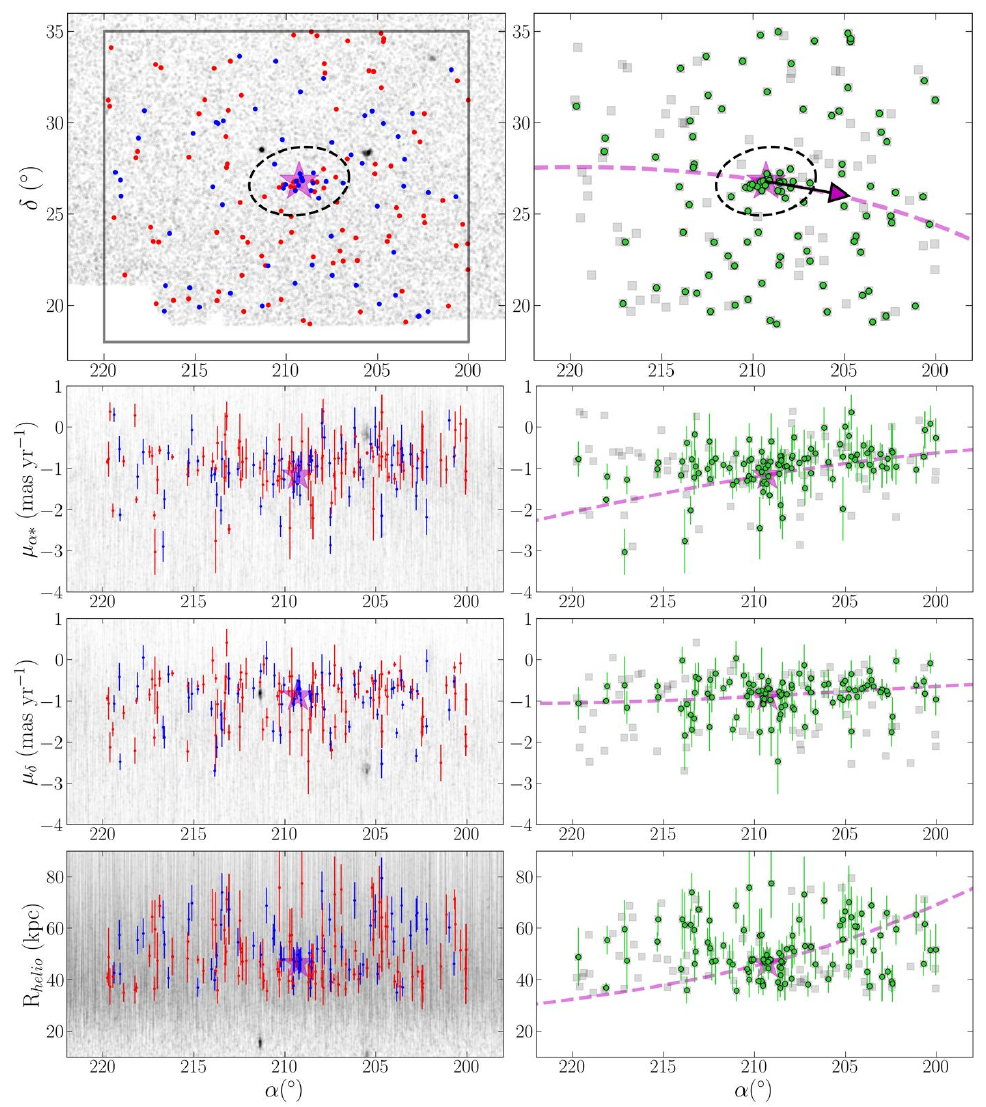}
    \caption{Equatorial positions, proper motions, and heliocentric distances of the BHB + RGB sample. The left panels show the subset of data used in the routine (BHBs in blue, RGBs in red) compared to the total catalogues in grey. The right panels showcase the sample surviving the routine (now as green points) against the original BHB + RGB subsample (now in grey squares). We also show the expected orbit of Boo3 in the magenta dashed tracks on the sky, in proper motions and distances, for direct comparison to the data. Boo3's systemic parameters are indicated as the magenta star, and we indicate 5$r_{h}$ as the black dashed ellipse. While the routine identifies the main body of Boo3 successfully, there is no clear evidence for any extended features such as a stellar stream.}
    \label{fig:UNIONS_sigmaclip}
\end{figure*}

Applying this method to the sample relies heavily on each star's proper motions and their errors. Given that the distances of this data are intentionally chosen to be rather large (35 $-$ 80 kpc), our sample’s proper motion errors overall tend to also be large (e.g., nearly 20\% of our sample have have percent errors $>$100\%). When we initially run the routine, we find that the fits are dominated by the closest stars in the sample due to their generally lower errors. This is evidenced by two factors in the weighted trendlines: (i) the direction of the trends are in complete opposition (i.e., perpendicular) to what we anticipate the direction should be, given the proper motions of Boo3, and (ii) the fit does not pass through the position of Boo3 in ($\alpha$, $\mu_{\alpha*}$). In order to ensure that the trendlines indeed remain consistent to that of Boo3 in the presence of substantial foreground contamination and high proper motion errors, we stabilize the fit by generating 10,000 MC sampled ``stars'' to represent the dwarf. We randomly sample the positions of mock data within 3$r_{h}$ and assign them reasonable proper motions, sampled under a Gaussian (whose mean and $\sigma$ are taken from Boo3's parameters in Table~\ref{tab:Boo3Params}). These mock stars are appended to our sample prior to running the routine, which serves to anchor the fit to Boo3's observables.
% sampled under a Gaussian 1-$\sigma$ uncertainties of Boo3 (see Table~\ref{tab:Boo3Params})

% (based on observations of Boo3) while allowing us to make an independent detection of Boo3’s debris. 

The subsample of surviving UNIONS RGBs and BHBs totals 166 sources, and after the sigma-clipping routine, our sample is limited to 103 stars across the $\sim$20$^{\circ}\times\sim$$20^{\circ}$ field. Figure~\ref{fig:UNIONS_sigmaclip} presents the positions and kinematics of the subsample (colored red and blue in the left panels; represented as grey squares in the right) prior to applying the routine, while green data in the right panels indicate the remaining sample.

Encouragingly, we observe an apparent grouping of stars coinciding with the position of Boo3 in the upper right panel of Figure \ref{fig:UNIONS_sigmaclip} that survive the sigma-clipping routine. These, upon further inspection, all have consistent proper motions and distances to that of Boo3 and are highly likely members, as they are located within 5$r_{h}$. However, we note that the other stars surviving the sigma-clipping routine have a nearly isotropic spatial distribution, which does not appear to favor a coherent stream-like structure. We note in the proper motion panels of Figure \ref{fig:UNIONS_sigmaclip} that, while these surviving stars indeed are approximately within to 3-$\sigma$ to the orbit (shown as the dashed pink lines), we conclude that many of this sample survives due to their large proper motion errors. The bottom panel of this figure further indicates this, as a substantial number of stars that survive the routine are inconsistent with the heliocentric distances predicted by the orbit. We note that distances were excluded from the sigma-clipping routine, and as such can be used in this way to confirm the presence of an underlying substructure in the remaining data.

\begin{figure}
    \centering
    \includegraphics[width=\linewidth]{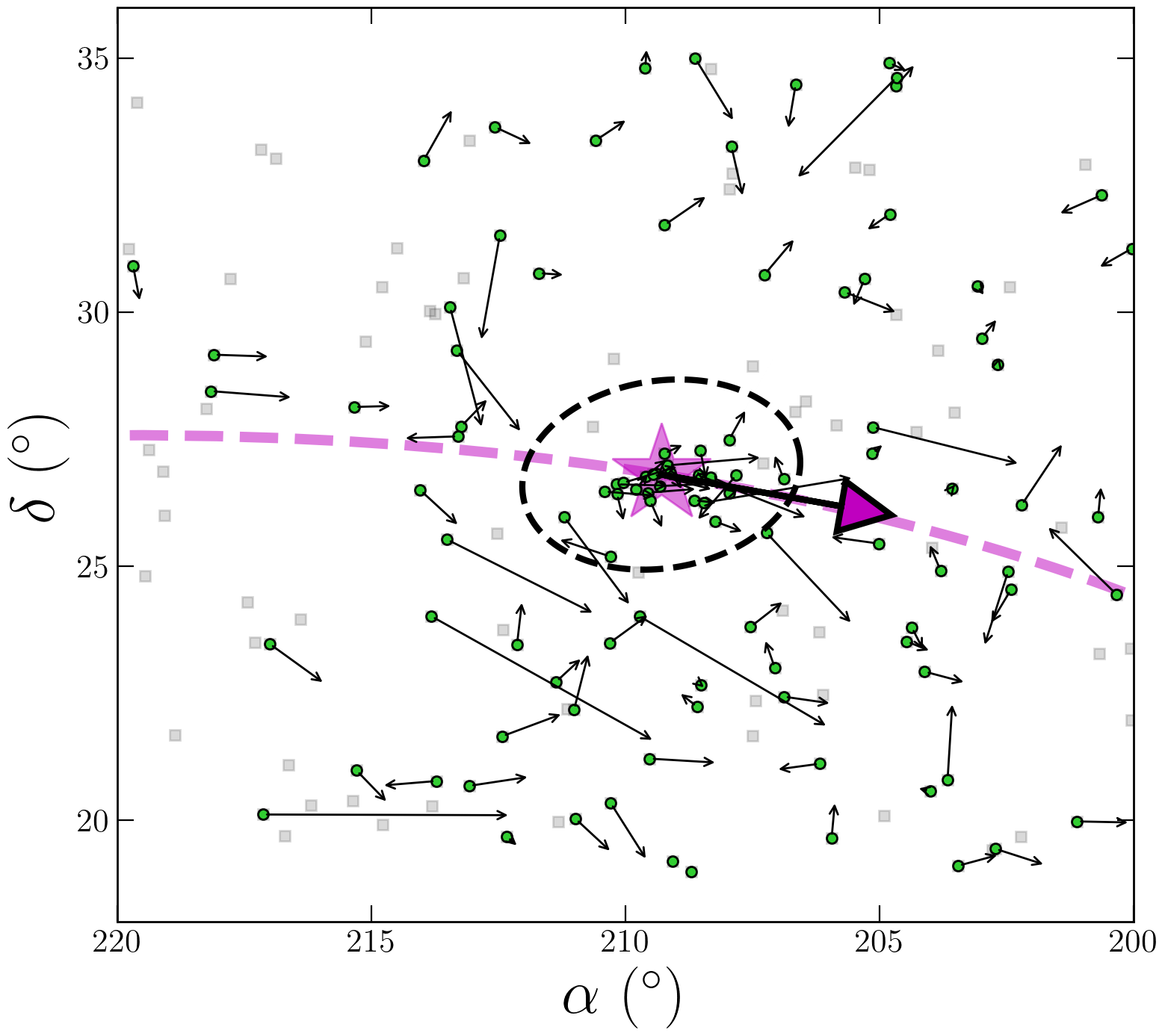}
    \caption{On-sky positions of the UNIONS BHB + RGB subsample (grey squares) compared to the sample remaining after the sigma-clipping routine (green). Vectors on these data represent the solar-corrected proper motions. The black dashed ellipse indicates 5$r_{h}$, for comparison. Stars stripped from Boo3 should have similar proper motions to that of the dwarf (magenta arrow), yet we only find a limited number of stars moving in the same direction.}
    \label{fig:UNIONS_sigclip_pmvect}
\end{figure}

To clean the sample of potential outliers, and determine whether these remaining stars are actually consistent with membership to Boo3, we conduct a brief sanity check. For every star surviving the sigma-clipping routine, we calculate its tangential motion (assuming distances from the photometric estimates in \citetalias{thomas2018} and \citetalias{thomas2019}) and plot the star and its motion vector on the sky. Note that the true tangential motion is essentially the star's proper motion vector, corrected for solar-reflex motion (which relies on each star's proper motion and photometric distance). If the star's vector appears to be moving in the same direction as the satellite, and its distance is also consistent to that of the orbit at that location on the sky, then we may consider it a star originating from Boo3.

Figure \ref{fig:UNIONS_sigclip_pmvect} shows the on-sky positions and corrected proper motion vectors (multiplied by a factor of 2 for better visibility) of the remaining stars in the sample. We also highlight Boo3’s solar-corrected proper motion vector in magenta, centered on the position of the dwarf and scaled by $\times$10. As shown in the figure, very few stars outside of 5$r_{h}$ have motions consistent with Boo3’s orbit; these are a small collection of 3 or so, at approximately at (RA, Dec) = (216$^{\circ}$, +29$^{\circ}$). Upon further inspection, we note that the distances to these stars are substantially different than the distances predicted by the orbit, and so these are not likely stellar debris of Boo3.

\subsection{Matched Filter Analysis}
\label{section:outer_MF}

In this section, we aim to continue our search for a stream in the vicinity of Boo3 using the matched filter (MF) technique described in \citet{grillmair2009}. This method, which is optimized to search for substructures over large areas of sky, ultimately constructs a map where stellar streams and satellites can be identified by eye. MW substructures appear as over-dense regions of the map, where the density of these objects are higher than the background. Matched filters have shown success in many works to detect numerous MW stellar streams (e.g., \citealt{rockosi2002, belokurov2006_FOS, shipp2018, ferguson2015, thomas2020}) including the work conducted in \citet{grillmair2009}. We begin by attempting to replicate the results of \citet{grillmair2009} in SDSS to re-examine the detection of Styx and Boo3. Subsequently, we also test this detection with the MF applied to the (slightly deeper) DELVE catalogue. Using both photometric surveys with the same detection method provides a sanity check that any faint or ambiguous features identified the MF maps are not spurious detections or due to observational artifacts.

Here, we briefly summarize the similarities in method used in \citet{grillmair2009} and this present work. Essentially, the MF constructs a weighted density map in which each star is assigned a likelihood of membership to an assumed stellar population (in our case, representative of Boo3 itself), according to its consistency in color and magnitude. The map is then smoothed to remove small-scale fluctuations, and later enhanced by subtracting a modeled background density. This method successfully improves the detection of satellites and streams by increasing the contrast between underlying structures and the MW foreground.

In our implementation, we calculate the weights to each star from a likelihood function that is modeled after both the satellite and a background CMD. These functions describe a star’s likelihood of association to Boo3 or the MW foreground based on its color and magnitude. We choose to represent these functions as ``lookup maps'' $-$ similar to the method described in \citet{McVenn2020} and \citet{jensen2024} $-$ wherein one can simply ``look up'' an individual star’s likelihood by locating its color and magnitude on this map. Each pixel in the lookup maps is 0.025 mag wide in color and magnitude.

The satellite CMD map is modeled after an old (12 Gyr) Padova isochrone and luminosity function (\citealt{girardi2002}, converted to \textit{Gaia} magnitudes using the corrections from \citealt{weiler2018}) assuming a metallicity of [Fe/H] = $-$2.1. The likelihood of membership on the CMD is derived from a Gaussian probability density function (PDF), whose mean is centered on the isochrone and whose width extends in color for a given bin in $g$. The width and amplitude of this Gaussian accounts for (i) the uncertainties of Boo3’s distance modulus, (ii) photometric errors as a function of magnitude, and (iii) the relative number of expected stars in a given population along the isochrone. To construct the satellite likelihood map, we first take a random sample of Boo3’s distance modulus within 1-$\sigma$ (see Table \ref{tab:Boo3Params}) and shift the isochrone to this magnitude. We then account for the distribution of stars anticipated per evolutionary phase by providing weights to the amplitudes, according to a Kroupa initial mass function (\citealt{kroupa2001}). The $\sigma$ of the Gaussian PDFs are essentially the photometric errors as a function of $g$ added in quadrature to a fixed intrinsic width of the RGB in color (0.1 mag). For 1000 individual Monte Carlo samples of the distance modulus, we repeat this process to create a series of 1000 likelihood maps. These maps are then summed together and normalized such that the final CMD satellite map integrates to unity. 

\begin{figure}
    \centering
    \includegraphics[width=1\linewidth]{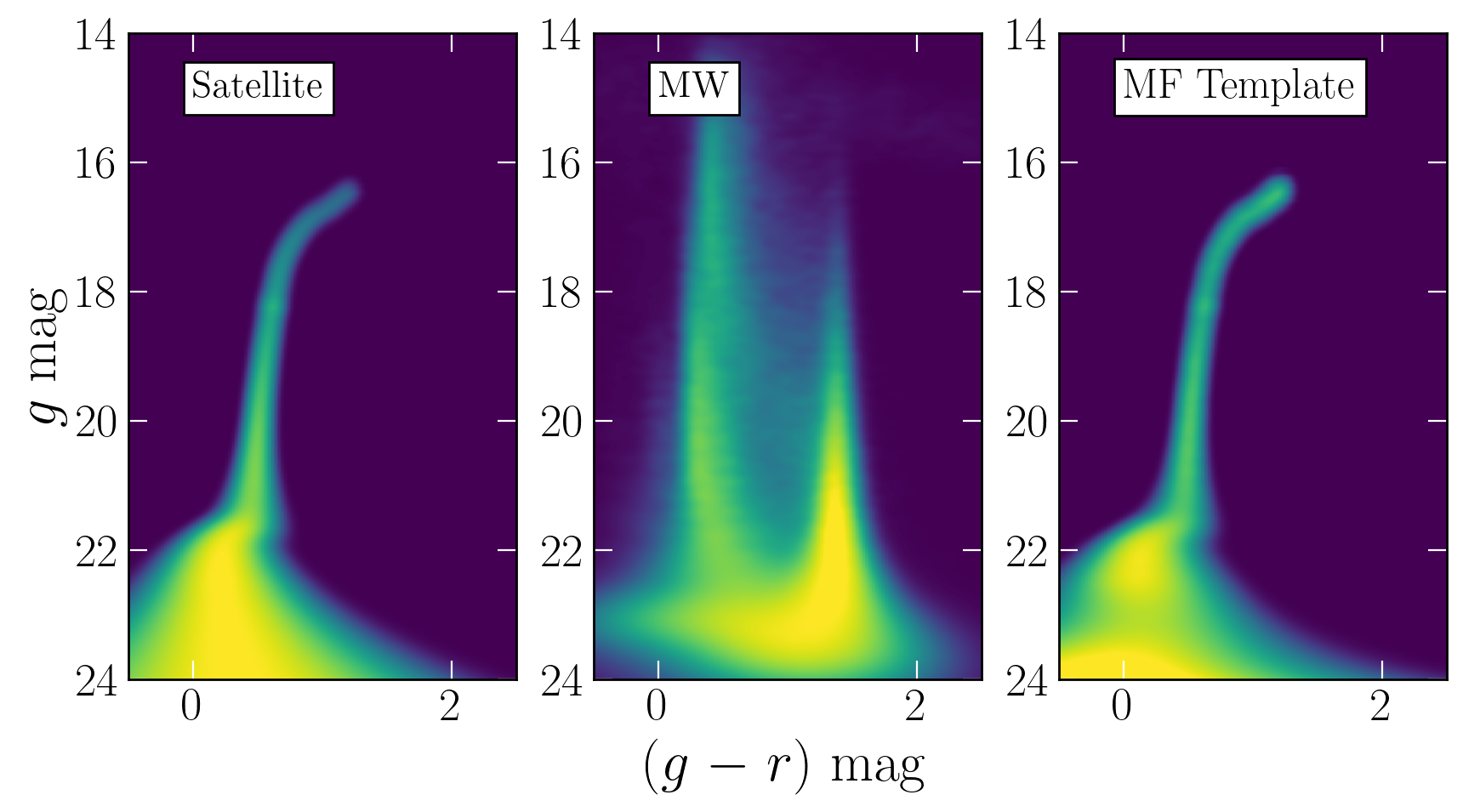}
    \caption{Example of the CMD ``lookup maps'' used to estimate a star's likelihood of association to Boo3. The final likelihood function in the right panel represents the matched filter template used to create the density map.}
    \label{fig:MF_template}
\end{figure}

In contrast, the foreground CMD likelihood map is constructed entirely empirically. To develop this lookup map, we select a region of sky distant to Boo3 that does not contain large-scale known substructure (e.g., away from the Sagittarius stream, seen better in Appendix Figure \ref{fig:MF_full}). We selected stars located between $214^{\circ}\leq\text{RA}\leq225^{\circ}$ and $+7^{\circ}\leq\text{Dec}\leq+16^{\circ}$ to represent the background, and their colors and magnitudes are then binned into a similar CMD pixel grid as the satellite CMD likelihood map. We then apply a bivariate Gaussian kernel to account for individual photometric uncertainties. The final background CMD likelihood map is then similarly normalized such that it integrates to unity.  

Finally, the total CMD likelihood map (i.e., the MF template) is constructed by dividing the satellite CMD map by that of the background. We show in Figure~\ref{fig:MF_template} examples of these lookup maps, where the right panel indicates the total CMD likelihood map used in this work. 

To create the matched filter density map, we compute the weighted counts by looking up each star’s likelihood in the MF template and summing the total counts per 0.1$^{\circ}$ pixel in tangent plane coordinates. For both surveys, we compute the matched filter map using $g$ vs ($g - r$) color-magnitudes. In the case of SDSS, we additionally benefit from the continuous coverage of the $i$-band. As conducted in \citet{grillmair2009}, we make a second matched filter map using $g$ vs ($g - i$) and sum these two maps together to make the final image. We note that, given the current survey footprint, it is not possible to take this second step using DELVE as we lack uniform $i$-band coverage this far north. Each band of each catalogue is limited in magnitude by the 5-$\sigma$ point source depths discussed in Section \ref{section:surveys}. We find that the SDSS and DELVE maps appear largely the same with these limits. As such, we attempt to enhance the contrast of the MF maps by taking a final broad limit of 15~$\leq$~$g$~$\leq$~22.5 mag (as was also done in \citealt{grillmair2009}) for both SDSS and DELVE.

\begin{figure*}
    \centering
    \includegraphics[width=1\textwidth]{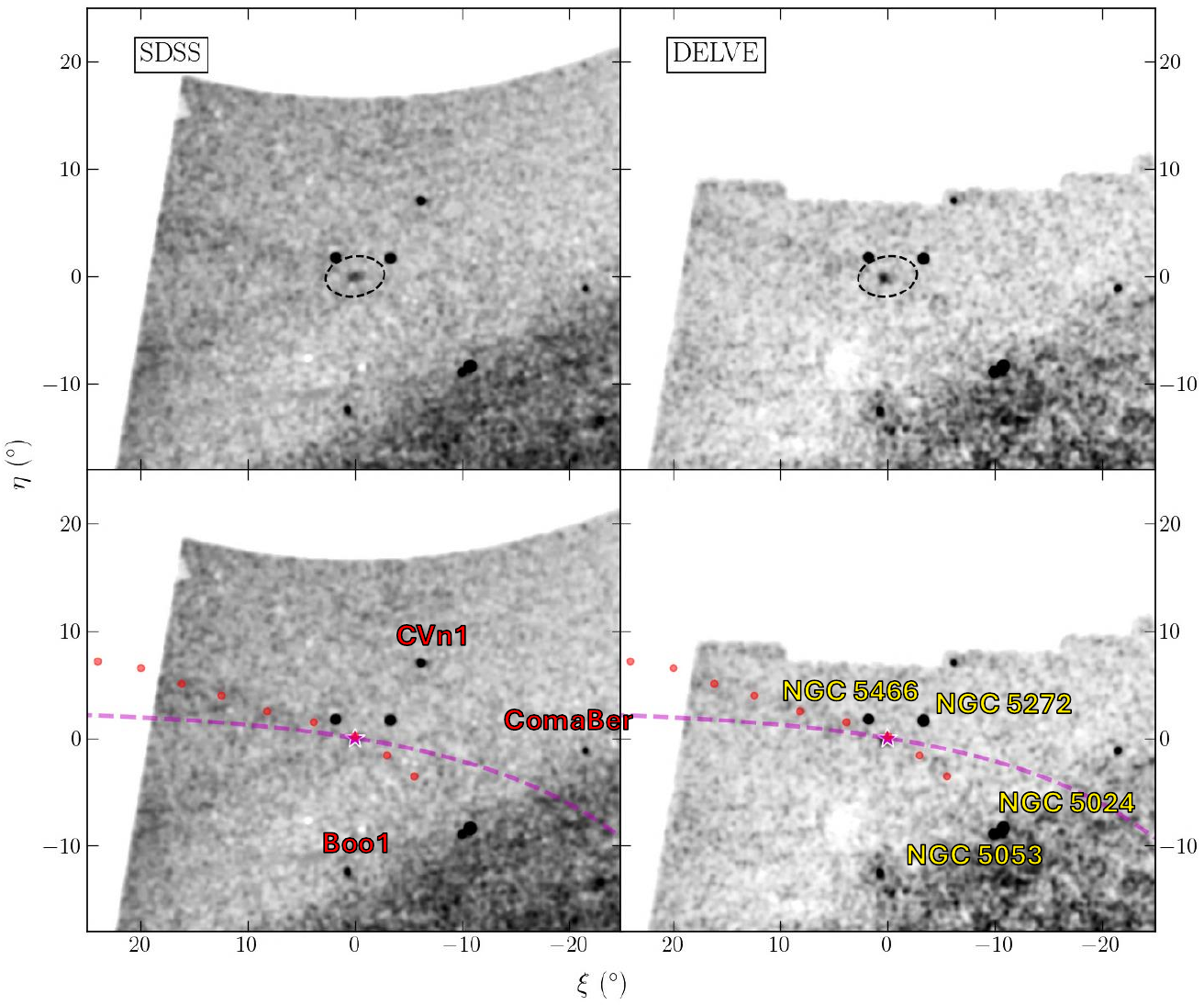}
    \caption{Tangent plane projections of the matched filter maps produced using SDSS (left panels) and DELVE (right panels). Note that the top and bottom panels are the same maps, just with added annotations. Boo3, located at (0,~0), is highlighted with the magenta star and black dashed ellipse (representing 5$r_h$). To indicate the differences between Boo3's trajectory and the approximate location of Styx, we plot both the expected orbit of Boo3 (magenta dashed line) and a range of nodes along the position of Styx (red points). Known satellites (dwarf galaxies and globular clusters) observed in these maps are also annotated.}
    \label{fig:MF_map_tangent}
\end{figure*}

% Note that these nodes, originally in SDSS survey coordinates, have been converted to equatorial and finally to the tangent plane. 

We then remove smaller-scale density fluctuations in the image by smoothing with a Gaussian kernel of 0.2$^{\circ}$. As a final step to enhance contrast, we subtract a background density map modeled after a 5$^{th}$-degree polynomial surface (after masking known globular clusters and dwarf galaxies). The final matched filter maps for both SDSS (left) and DELVE (right) in a tangent plane projection are presented in Figure~\ref{fig:MF_map_tangent}.

As shown in all panels of Figure~\ref{fig:MF_map_tangent}, we successfully recover all previously reported globular clusters and dwarf galaxies (labeled in the bottom right and left panels, respectively), including the faint and diffuse Boo3 dwarf centered at the origin. We additionally recover a portion of the Sagittarius stream as a large overdensity in the bottom right corner of every panel (see also Figure \ref{fig:MF_full} in the Appendix). Boo3 is clearly identifiable as a small overdensity in both catalogues, indicated with a dashed ellipse (representing 5$r_h$) or a magenta star. 

The matched filter maps indeed look very similar in both SDSS and DELVE, with some minor differences in sky coverage and overall footprint. In the SDSS panels, we do observe a small streak to the lower left (southeast), but this feature is not observed in the accompanying DELVE map. We argue that this feature is indeed an artifact of the SDSS observing strategy, which surveys the sky in striping patterns along great circles. We demonstrate this argument further in Figure \ref{fig:SDSS_streaking} in the Appendix, where stars down to magnitudes of $g$~=~24 are plotted with the matched filter technique. Given the streak around Boo3 in both the absolute density (left) and the MF (right), it appears that the small-scale fluctuations produced by the observation patterns may in fact grow in size and sum together once smoothed by the Gaussian kernel. We also explore whether detection of Boo3's debris may be obscured by dust in Appendix Figure \ref{fig:extinction}, but the extinction values do not appear to be large enough to explain this substructure. 

% Even if this substructure were real, we note that because it lies within 5$r_{h}$ of the dwarf that it hardly constitutes a lengthy tidal stream. 

% and furthermore is perpendicular to the orbit trajectory and putative path of Styx (discussed further in Section \ref{section:disc})

In both maps, however, we do observe a faint feature to the upper left (northeast) of Boo3. To compare to the reported position of Styx to this structure, we plot a series of red nodes in the bottom panels of Figure \ref{fig:MF_map_tangent} representing its approximate location on the tangent plane. This was done by (i) digitizing a series of nodes in the original MF maps approximately centered on the stream and recording their positions in SDSS survey coordinates ($\lambda$,~$\eta$)$_{SDSS}$, (ii) manually converting their positions to equatorial coordinates using the transforms provided in \citet{stoughton2002_SDSScoords} (equations are reproduced in Appendix \ref{chapter:appendix}), and (iii) applying a final conversion from ($\alpha$, $\delta$) to the tangent plane coordinates ($\xi$, $\eta$) where the origin is centered on Boo3. Indeed, we find that these nodes appear to correspond to the position of the Styx stellar stream. 

\begin{figure*}
    \centering
    \includegraphics[width=\textwidth]{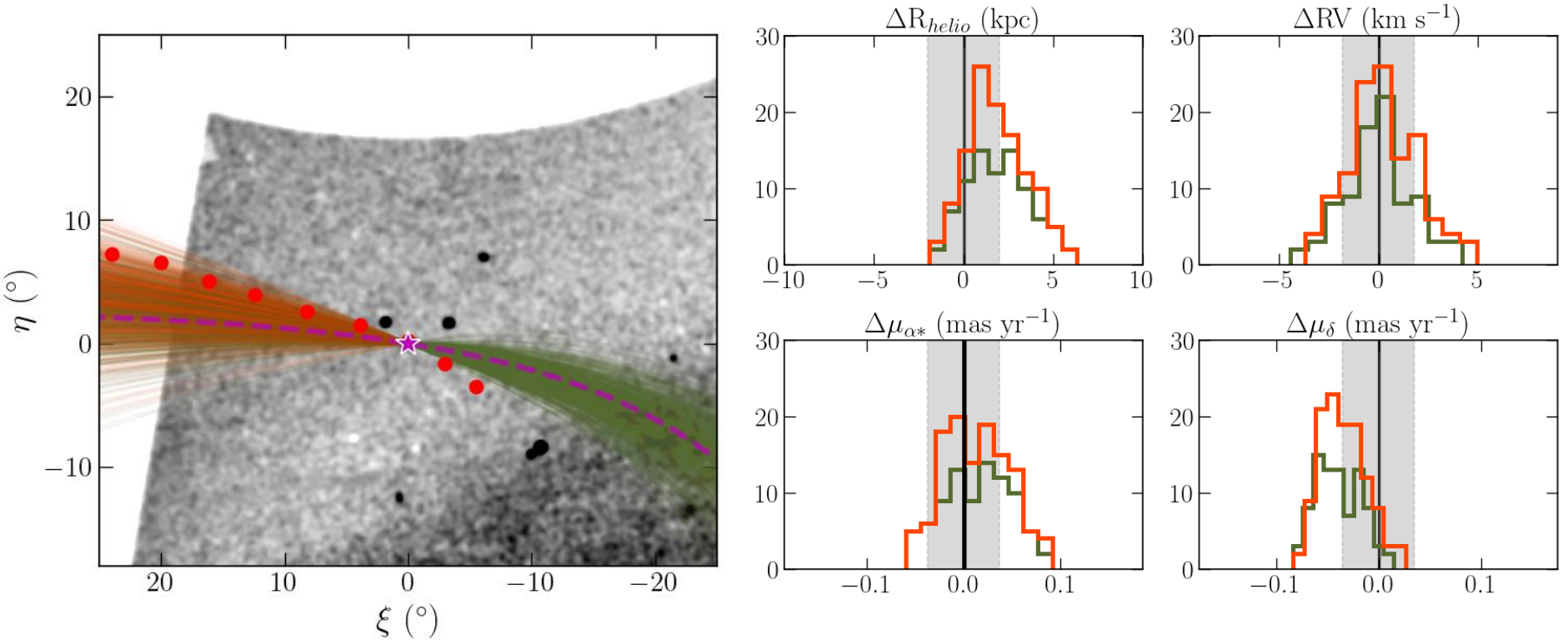}
    \caption{MC sampled orbits of Boo3 in the static (olive) and time-evolving (orange). \textbf{Left:} As shown in the tangent plane, the Styx stream (red nodes) does not obviously align with a majority of the explored orbits. The location of Boo3 and its mean orbit (for both potentials) are indicated in magenta. \textbf{Right:} Distribution of initial conditions used, for only the orbits which fall within 0.5$^{\circ}$ of at least 3 nodes (including the node located at Boo3). Each panel represents the difference between an initial condition and the mean observable of Boo3, and we additionally highlight the 1-$\sigma$ error for each variable in grey. Note that all initial conditions align with Boo3's mean observables within 1-$\sigma$, though the majority of proper motions in declination ($\mu_{\delta}$) lie outside this.}
    \label{fig:MC_orbits}
\end{figure*}

While we do find consistency in that these nodes do intersect with Boo3, they do not appear to follow the trajectory of the satellite given by its orbit (shown as the dashed magenta line). To confirm this inconsistency, we took 1000 MC samples of Boo3's initial conditions (proper motions, distance, and radial velocity), integrated these orbits in the static and time-evolving potentials, and compared Styx's position to these orbits. Figure~\ref{fig:MC_orbits} presents these results, where we observe that only a few extreme cases appear to correspond with the position of Styx. In fact, we find that only 6\% of orbits in the evolving potential and 4\% in the static potential intersect within 0.5$^{\circ}$ of at least 3 nodes, including the node located at Boo3. Of these corresponding orbits, we also determine the distribution of their initial conditions (subtracted by the mean parameter of Boo3) shown as histograms in the right panel. It is clear from this figure that all MC sampled parameters producing these extreme orbits fall at or close to 1-$\sigma$, except for $\mu_{\delta}$ whose distribution falls largely outside of the typical range of values. Indeed, only 1 out of 1000 of the MC sampled static potential orbits align with the nodes in the leading arm.

Given that so few orbits correspond to the position of Styx, and that we do not identify any debris in the stellar tracer catalogues beyond 5$r_{h}$ that are consistent with Boo3's orbit, we are unable to claim that Styx and Boo3 are indeed linked structures. In the following section, we discuss implications for these results when examining the dynamical properties derived for Boo3.

\section{Discussion}
\label{section:disc}

In this work, we conducted wide-field searches for stream debris by exploring both (i) the UNIONS BHB/RGB stellar tracer catalogues for samples of co-moving stars consistent to stream membership, and (ii) wide-field photometric surveys in search of stellar overdensities that are aligned with Boo3's orbit. We conclude that we are unable to confirm the detection of \textit{any} extended debris of Boo3 $>$5$r_{h}$. Our null results appear in conflict with previous findings regarding the satellite: it is well established that Boo3 is an extremely diffuse and sizable dwarf, whose large velocity dispersion and non-uniform morphology implies it is a system that is \textit{not} in dynamic equilibrium. Even in our current work, we find that Boo3 may be strongly affected by MW tides given the proximity of its 2 $-$ 3 most recent pericenters ($<$20~kpc, which is argued to be enough to disrupt a typical dwarf; see N-body simulations and relevant arguments by \citealt{read2006}). We further find that its orbit is very eccentric (0.88) and polar (yet slightly retrograde), which may additionally play a role in the level of observed disruption. Why then, are we unable to observe any tidal tails of Boo3?

In the following subsections, we first examine the distribution of a putative stream in the particle-spray models previously described in Section \ref{section:orb_partspray}. We then posit a few dynamical interpretations justified by the literature, involving a potential interaction with the Galactic bar and Boo3's very eccentric orbit.

% After probing multiple stellar tracer catalogues for samples of co-moving stars and photometric surveys in search of true substructures presenting themselves as stellar overdensities, we conclude that we are unable to confirm the detection of \textit{any} extended debris of Boo3 $>$5$r_{h}$. 

\subsection{A Missing Link of Stream Debris}

\begin{figure*}
    \centering
    \includegraphics[width=1\textwidth]{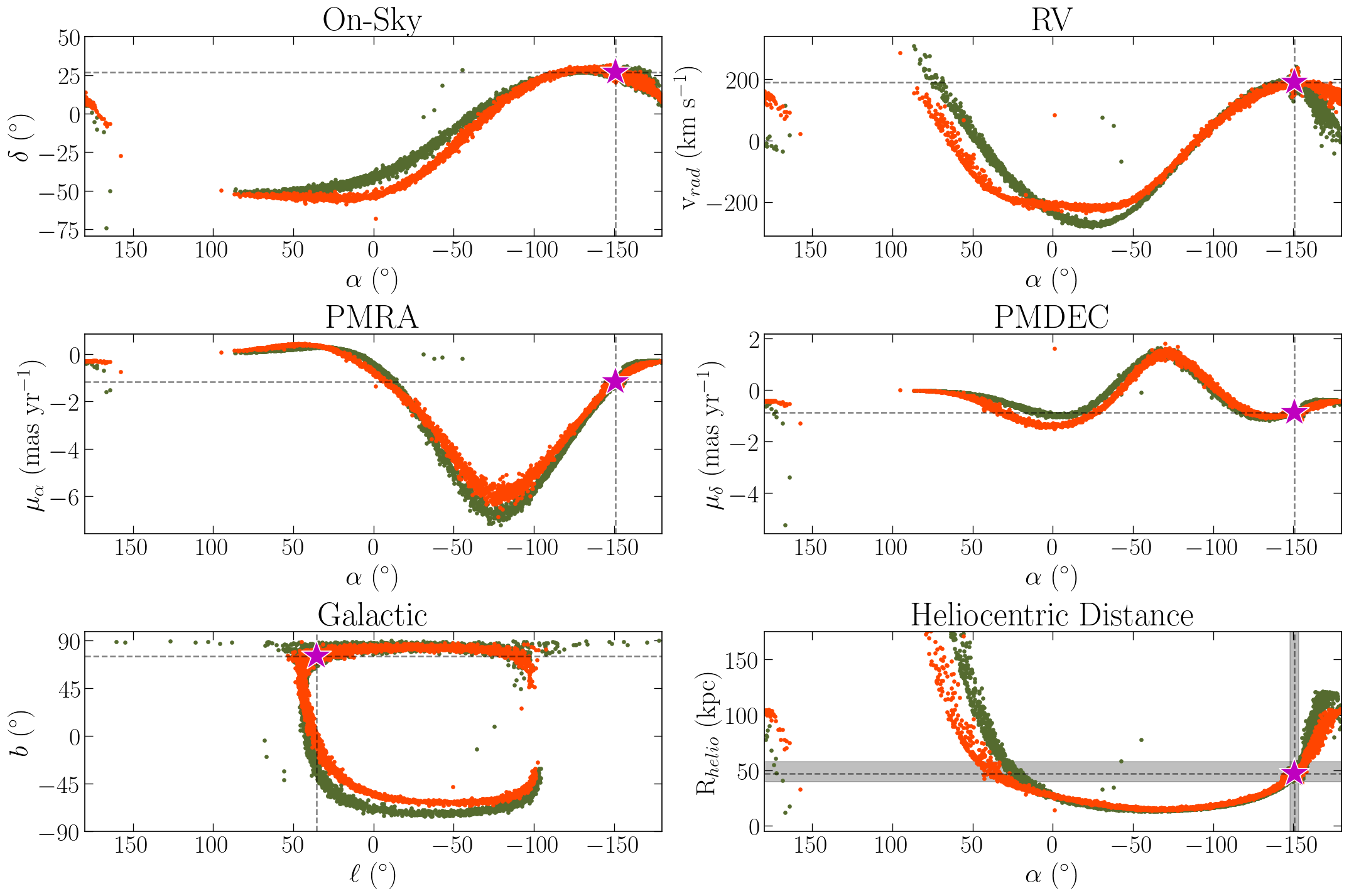}
    \caption{Distribution of simulated particles in various observables versus their on-sky position in RA ($\alpha$) or Galactic longitude ($\ell$). The phase-space trends of sprayed particles for both the static (olive) and evolving (orange) potentials are largely similar, including the significant heliocentric gradient of the leading arm ($-$180$^{\circ}$~$\lesssim$~$\alpha$~$\lesssim$~$-$150$^{\circ}$). For comparison, we indicate the location of Boo3 in each observable as the magenta star, further highlighted by the grey dashed cross-hairs.}
    \label{fig:particle_spray}
\end{figure*}

Some initial insight into where Boo3's stream debris may be located, can be interpreted from the phase-space distribution of our particle-spray model. In Figure~\ref{fig:particle_spray}, we plot the locations of Boo3’s putative debris in various on-sky observables (equatorial and Galactic coordinates, proper motions, heliocentric velocities, and distances) versus their on-sky positions in $\alpha$ and color-coded by the host potential. For comparison, we highlight the present-day observables of Boo3 with a magenta star and grey dashed crosshairs. As observed in each panel, the overall trends between the static and evolving potentials are largely similar with minor offsets in the trailing arm. The continuity of these trends, which are similar in \textit{both} potentials, imply that we \textit{should} have been able to observe a coherent stream structure if indeed there is a continuous distribution of Boo3 debris. Certainly, it is logical to assume we \textit{should} have seen debris closest to the system itself, in either of the UNIONS-identified RGB or BHB catalogues, if Boo3 indeed has tidal tails.

On closer inspection however, Boo3’s debris is predicted to exhibit a steep distance gradient as shown in the bottom right panel of Figure \ref{fig:particle_spray}. To showcase the extent of this gradient, we highlight the range of distances spanning area on the sky equal to 5$r_{h}$ ($\pm$2.5$^{\circ}$; slightly larger than the CaHK footprint of Figure~\ref{fig:Pr_Gaia_Mems}) around Boo3 in grey vertical and horizontal bands. As seen in the heliocentric panel, both particle-spray models span distances of $\sim$39~kpc in the trailing arm and out to $\sim$60~kpc in the leading arm within 5$r_{h}$. In just the span of 5$^{\circ}$ in RA, Boo3’s putative debris may be experiencing a substantial distance gradient such that our view of the stream may be obscured, and it may not be possible to observe a clearly extended structure. We note that the distance moduli corresponding to the grey bands in Figure \ref{fig:particle_spray} are actually within the 1-$\sigma$ errors of Boo3's estimated distance modulus. This means that, even though the orbit shows a large distance gradient in the vicinity of the dwarf itself, the likelihood maps used to construct the MF would include stars at these distances. Thus, if there was indeed a structure located here, it would still have been detected as an overdensity in the MF maps.

A truncated view of the leading arm due to this distance gradient is certainly a feasible reason as to the lack of observed Boo3 debris. However, we note that the heliocentric distance trends of the trailing arm does suggest that there may be a portion of Boo3’s debris that is orbiting at much closer distances. The closest extension, ranging from 10 $-$ 20 kpc, is predicted to occupy a large area of the sky from $245^{\circ}\lesssim\text{RA}\lesssim340^{\circ}$ and $-47^{\circ}\lesssim\text{Dec}\lesssim+25^{\circ}$. In an attempt to link any previously detected streams to Boo3, we also searched the MW stellar streams compilation in \texttt{galstreams} (\citealt{mateu2023}) for any reported structures that broadly match the same positions and distances as this fragment. Unfortunately, no such feature in this area of sky has yet been catalogued. Given the relatively nearby proximity of Boo3’s trailing debris, it is indeed surprising that no structure in this vicinity has yet been found.

\subsection{Potential Dynamical Interpretations}

Another possible interpretation for the absence of Boo3’s tidal tails could be explained by an interaction with the Galactic bar. Recent studies have demonstrated that bar-induced perturbations on stream stars can affect the stream's overall morphology, particularly for systems with pericenters $\lesssim$10~kpc (e.g., \citealt{thomas2023, bonaca_price-whelan2025}). For example, simulations by \citet{pearson2017} found that the dispersal of stars in the Pal 5 stream may be explained by torques imparted by the bar at pericenter. In this scenario, \citet{pearson2017} argue that the additional torque (either positive or negative, depending on the bar’s orientation) increases/decreases the star’s angular momentum and therefore also its orbital energy. If these perturbations \textit{add} torque, L$_{z}$ increases, total E becomes less negative, and star particles migrate to orbits at larger radii and orbit the MW at slower speeds. The inverse is true if the added torque is negative: the star’s orbit becomes shorter and its velocity increases to be pulled ahead. Thus the influence of the Galactic bar can result in the dispersal of stream stars, vastly decreasing the overall density along the stream. 

Other studies of the Ophiuchus stream (\citealt{bernard2014}) have shown that bar-driven morphological changes could explain its interesting structure, such as shortening/lengthening of its tails due to bar resonances (``shepherding''; \citealt{hattori2016}) or dispersal of its stream stars due to the progenitor’s chaotic orbit (``chaotic fanning''; \citealt{price-whelan2016}). In the latter scenario, \citet{price-whelan2016} argue that the bar’s influence may force disrupted stream stars into chaotic regions of E and L$_{z}$. This process results in a dramatic drop in stream density, such that only the most recently disrupted material will be detected as a coherent structure. Given the simplicity of our current models in which we do not include a rotating central bar, we are currently unable to test if chaotic fanning could affect Boo3's debris, but recommend this investigation for future works. 

%%%%ADD part about testing for chaotic orbit ??

In the case of the Boo3 dwarf, we find that the satellite has likely experienced a close enough pericenter such that the bar’s influence may be considered non-negligible. For both the static and evolving potentials, we find that Boo3’s closest pericenter is $\sim$8~kpc (congruent to findings from \citealt{battaglia2022} and \citealt{pace2022}). However, we also argue that this most recent pericenter ($\sim$140 Myrs ago) is the \textit{only} instance in Boo3’s orbit where its stream may have been affected by the bar. We find that, in the past 5 Gyrs of its orbit, the pericenters of Boo3 are nearly \textit{double} this value (on the order of $\sim$16~kpc). We therefore suggest that, \textit{if} the bar has had any interaction with Boo3’s stream, it \textit{must} have occurred only recently. Prior to this incident, we argue that the putative stream was likely unaffected by the bar. 

A perhaps simpler interpretation of Boo3's missing tails could be derived from its orbital properties overall. N-body simulations of dwarf spheroidal galaxies in \citet{penarrubia2008} show that a satellite's stars can be removed essentially impulsively at pericenter for systems on highly eccentric orbits. At pericenter, the tidal tails form quickly, leaving the bound core almost immediately after the encounter. By the time the dwarf reaches apocenter, the remnant may show essentially no sign of perturbation in its outskirts (see Figure 2 in \citealt{penarrubia2008}). It may be the case that Boo3's tails are difficult to detect due to its highly eccentric orbit (0.88). However, we note that Boo3 left pericenter only recently, and indeed will not reach apocenter for another $\sim$1.2 Gyrs. Unfortunately, our particle-spray models are limited in reproducing these sorts of trends, as the massless ejected particles do not capture the appearance of rapid mass loss at pericenter. At this time, we leave a more sophisticated analysis of Boo3’s disruption to future works. 

% If this debris was recently stripped, we assume that we would expect to have identified it in the UNIONS stellar tracer catalogues, which were some $\sim$20$^{\circ}\times\sim$20$^{\circ}$ of sky. 

% Indeed, we attempted to search for any debris of Boo3 at large distances and did not find any clear evidence of stellar debris within the UNIONS stellar tracer catalogues. 

%%%%%%%%%%%%%%%%%%%%%%%%%%%%%%%%%%%%%%%%%%%%%%%%%%%%%%%%%%%%%%
\section{Conclusions \& Summary}
\label{section:summary}
% Boo3 is a particularly enigmatic system, whose observational and orbital properties strongly suggest it is a dwarf galaxy (given its size and apparent magnitude) in the throes of complete disruption. 

In this work, we carried out a comprehensive investigation to the dynamical state of the ultra-faint dwarf galaxy, Boo3, which has been previously reported to be in the throes of complete disruption. After identifying extended stellar candidates of Boo3 out to $\sim$14$r_{h}$ in \textit{Gaia}, we sought new narrow-band CaHK imaging to derive photometric metallicities for 39 RGBs out to 5$r_{h}$. These observations serve to better constrain the dwarf's metallicity and metallicity dispersion, which we independently report as [Fe/H] = $-$2.1 $\pm$ 0.1 and $\sigma_{\rm Fe/H}$~=~0.2 $\pm$ 0.1, respectively. Encouragingly, we find both of these measurements fall within 1-$\sigma$ of more recent reportings (\citealt{li2026}). Given our newly established estimates for the metallicity and dispersion, we then introduced a secondary dataset to conduct a wider search for Boo3’s extended stellar debris: the stellar tracer catalogues from UNIONS. For these datasets, we applied a sigma-clipping routine to their spatial and kinematic (i.e., proper motions) properties in order to filter out stars whose properties broadly align with a putative stream. While our analysis did uncover a significant overdensity of co-moving stars within 5$r_{h}$, no coherent stream structure outside of this radius was observed in the data. Not only were the heliocentric distances of stars at $>$5$r_{h}$ found to be inconsistent to Boo3’s orbit, but their solar-corrected proper motion vectors also did not appear to align with the direction of the orbit, yielding further confirmation that these stars are inconsistent with membership with Boo3.
% , of which the mean agrees with other recent estimates within 1-$\sigma$ (\citealt{li2026}) but decreases the reported metallicity dispersion. 

% These observations led to robust measurements
% . These observations led to better constraints on Boo3's metallicity and metallicity dispersion, indicating that the syste

% By exploring Boo3's \textit{Gaia}-based candidates, we observed evidence for an extended stellar population 

% We obtained follow-up CaHK observations of the system, spanning the innermost regions out to $\sim$5$r_{h}$, to re-derive its metallicity and dispersion with substantially more stars. Our photometric metallicities constrain Boo3’s mean metallicity to be [Fe/H] = $-$2.1 $\pm$ 0.1 and its dispersion as $\sigma_{\rm Fe/H}$~=~0.2 $\pm$ 0.1. 

Finally, we conducted a wide-field search for Boo3’s stellar stream by reproducing the matched filter methodology used in the initial discovery paper of Boo3 and Styx, and applied this analysis independently to both SDSS and DELVE photometric catalogues. In both datasets, we recover many known globular clusters, dwarfs (including Boo3), and streams (Styx and Sagittarius). However, we do not find convincing evidence that Boo3 is affiliated to Styx, as Boo3's mean orbit is offset in comparison, and furthermore only a handful of orbits align with just a small portion ($\geq$3 nodes in Figure \ref{fig:MF_map_tangent}) of the Styx stream. This misalignment, in conjunction with the lack of stellar tracers that would indicate a coherent stream structure from Boo3, indicates that Styx may not be physically associated with Boo3. 

% Though we recover many known globular clusters and dwarfs (including Boo3) and Styx, in neither dataset do we recover convincing evidence for an extended stream connected to Boo3. Additionally, we found that Boo3’s orbit is offset compared to the putative location of Styx (derived from the original SDSS matched filter map). 

% A more detailed study of the dwarf’s debris will be more feasible with the completion of deep photometric surveys such as UNIONS, whose data (when observations are complete) will achieve similar depths to the first data release of the Rubin Observatory’s \textit{Legacy Survey of Space and Time} (\citealt{ivezic2019, bianco2022}). Complementary surveys like Euclid (\citealt{euclid_collab_2024}) will furthermore provide better star-galaxy separation, such that probing the main sequence of the stream will be much more accessible with upcoming surveys.

Our exploration of Boo3’s putative stream in phase-space using particle-spray models and its overall orbit indicated a few interesting reasons why the stream may not be currently observable. In particular, we found that the Boo3's stream may be experiencing a substantial distance gradient such that the leading arm may be heavily truncated. However, our models predict that the trailing arm debris should be located at relatively nearby heliocentric distances ($\sim$10 $-$ 20 kpc), yet we found no such structure yet reported in the literature. We argue that the absence of both the leading and trailing structures \textit{could} be explained by a past interaction with the Galactic bar. We note however, that given Boo3’s orbit, it is only within the most recent pericenter that a perturbation acting to erase the stream could have occurred. A simpler scenario may be that Boo3's very eccentric orbit may have caused debris to have been removed quickly at pericenter; however, having only experienced pericenter recently, we would expect that Boo3's debris would still be near enough to the system to be detected.

% If correct, Boo3 may represent a rare example of a disrupting dwarf galaxy whose debris morphology has been shaped not only by the MW halo, but also by the inner Galaxy’s rotating bar. 

Boo3 therefore remains an enigmatic system: while multiple lines of evidence suggest that it is likely experiencing ongoing tidal disruption, its stellar debris remains difficult to identify. If Boo3 is disrupting, our non-detection of coherent debris implies that the surface brightness of its tidal tails falls below current detection limits. Given the surface brightness of Boo3 and its detection significance above the background in the matched filter map, we take the ratio of the MF surface density within 1$r_{h}$ of Boo3 to that of a similar area along the orbit, and estimate an upper limit on the stream's surface brightness to be $\sim$33.4 mag arcsec$^{-2}$. Deeper photometric follow-up, paired with spectroscopic confirmation of membership and improved star-galaxy separation from space-based telescopes such as Euclid \citep{euclid_collab_2024}, should allow a better characterization of Boo3's putative stream. More detailed dynamical modeling of Boo3's disruption, including the influence of the LMC and a rotating Galactic bar, will also be essential for determining whether the stream has been strongly perturbed or dispersed. Indeed, such work is already underway (Boyea et al., in prep.). Studying Boo3's disruption offers not only an opportunity to better understand the tidal evolution of low-mass dwarf galaxies, but also a potential new probe of dynamical processes in the innermost regions of the MW.

\section*{Acknowledgments}
% \begin{acknowledgments}
% We thank all the people that have made this AASTeX what it is today.  This
% includes but not limited to Bob Hanisch, Chris Biemesderfer, Lee Brotzman,
% Pierre Landau, Arthur Ogawa, Maxim Markevitch, Alexey Vikhlinin and Amy
% Hendrickson. Also special thanks to David Hogg and Daniel Foreman-Mackey
% for the new "modern" style design. Considerable help was provided via bug
% reports and hacks from numerous people including Patricio Cubillos, Alex
% Drlica-Wagner, Sean Lake, Michele Bannister, Peter Williams, and Jonathan
% Gagne.

We thank Dr. Carl Grillmair for his incredibly influential work in the discovery and characterization of Milky Way stellar streams. We also thank Dr. Anke Ardern-Arentsen for her assistance in using and applying the photometric metallicity code from which our results were derived. 

AWM and JJ acknowledge support of a Natural Sciences and Engineering Research Council of Canada Discovery Grant RGPIN2018-03853. JJ also acknowledges support from the National Aeronautics and Space Administration (NASA) under grant No. 80NSSC25K0365. GFT acknowledges support from the Agencia Estatal de Investigaci\'on del Ministerio de Ciencia en Innovaci\'on (AEI-MCIN) under grant number PID2023-150319NB-C21 and the grant RYC2024-051016-I funded by MCIN/AEI/10.13039/501100011033 and by the European Social Fund Plus. RE acknowledges support from the National Science Foundation (NSF) grant
AST-2206046. Support for program JWST-AR-02352.001-A was provided by NASA through a grant from the Space Telescope Science Institute, which is operated by the Association of Universities for Research in Astronomy, Inc., under NASA contract NAS 5-03127. This material is based upon work supported by NASA under Grant/Agreement No. 80NSSC24K0084 as part of the Roman Large Wide Field Science program funded through ROSES call NNH22ZDA001N-ROMAN. NM ackowledges support by the International Space Science Institute (ISSI) in Bern, through ISSI International Team project 540 (The Early Milky Way). This work was supported by the Programme National Astro of CNRS/INSU with INP and IN2P3, co-funded by CEA and CNES. G.E.M. acknowledges financial support from Natural Sciences and Engineering Research Council of Canada (NSERC) through grant RGPIN-2022-04794 and from an Arts \& Science Postdoctoral Fellowship at the University of Toronto. 

The work detailed above was conducted at the University of Victoria in Victoria, British Columbia, as well as in the Township of Esquimalt in Greater Victoria. We acknowledge with respect the Lekwungen peoples on whose unceded traditional territory the university stands, and the Songhees, Esquimalt and WS\'{A}NE\'{C} peoples whose historical relationships with the land continue to this day. 

This paper includes data gathered with the 3.6-meter Canada-France-Hawaii Telescope (CFHT) located at Maunakea Observatories, Hawai'i. Observations obtained at CFHT are operated by the National Research Council of Canada, the Institut National des Sciences de l'Univers of the Centre National de la Recherche Scientifique of France, and the University of Hawai'i. CFHT is located on Maunakea on Hawai'i Island, a mountain of considerable cultural, natural, and ecological significance. Maunakea is a sacred site to Native Hawaiians, also known as K\=anaka \textquoteleft{}\=Oiwi. We would like to thank the CFHT Operations and Software Groups for their contributions and diligence in maintaining observatory operations; the CFHT Astronomy Group for their observation coordination and data acquisition efforts; and the CFHT Finance \& Administration Group for their contributions to the management and administration of the observatory. This work is also based on observations obtained with MegaPrime/MegaCam, a joint project of CFHT and CEA/DAPNIA.

% Kānaka ʻŌiwi

% This paper includes data gathered with the 3.6-meter Canada-France-Hawaii Telescope located at Las Campanas Observatory, Chile. 

This work has made use of data from the European Space Agency (ESA) mission $\textit{Gaia}$ (\url{https://www.cosmos.esa.int/gaia}), processed by the $\textit{Gaia}$ Data Processing and Analysis Consortium (DPAC, \url{https://www.cosmos.esa.int/web/gaia/dpac/consortium}). Funding for the DPAC has been provided by national institutions, in particular the institutions participating in the Gaia Multilateral Agreement.

We are honored and grateful for the opportunity of observing the Universe from Maunakea and Haleakala, which both have cultural, historical and natural significance in Hawaii. This work is based on data obtained as part of the Canada-France Imaging Survey, a CFHT large program of the National Research Council of Canada and the French Centre National de la Recherche Scientifique. Based on observations obtained with MegaPrime/MegaCam, a joint project of CFHT and CEA Saclay, at the Canada-France-Hawaii Telescope (CFHT) which is operated by the National Research Council (NRC) of Canada, the Institut National des Science de l’Univers (INSU) of the Centre National de la Recherche Scientifique (CNRS) of France, and the University of Hawaii. This research used the facilities of the Canadian Astronomy Data Centre operated by the National Research Council of Canada with the support of the Canadian Space Agency. This research is based in part on data collected at Subaru Telescope, which is operated by the National Astronomical Observatory of Japan.
Pan-STARRS is a project of the Institute for Astronomy of the University of Hawaii, and is supported by the NASA SSO Near Earth Observation Program under grants 80NSSC18K0971, NNX14AM74G, NNX12AR65G, NNX13AQ47G, NNX08AR22G, 80NSSC21K1572 and by the State of Hawaii.

Funding for the Sloan Digital Sky  Survey IV has been provided by the  Alfred P. Sloan Foundation, the U.S.  Department of Energy Office of  Science, and the Participating  Institutions. 

SDSS-IV acknowledges support and resources from the Center for High Performance Computing at the University of Utah. The SDSS website is \url{www.sdss4.org}.

SDSS-IV is managed by the Astrophysical Research Consortium for the Participating Institutions of the SDSS Collaboration including 
the Brazilian Participation Group, the Carnegie Institution for Science, Carnegie Mellon University, Center for Astrophysics | Harvard \& Smithsonian, the Chilean Participation Group, the French Participation Group, Instituto de Astrof\'isica de Canarias, The Johns Hopkins University, Kavli Institute for the Physics and Mathematics of the Universe (IPMU) / University of Tokyo, the Korean Participation Group, Lawrence Berkeley National Laboratory, Leibniz Institut f\"ur Astrophysik Potsdam (AIP),  Max-Planck-Institut f\"ur Astronomie (MPIA Heidelberg), Max-Planck-Institut f\"ur Astrophysik (MPA Garching), Max-Planck-Institut f\"ur Extraterrestrische Physik (MPE), National Astronomical Observatories of China, New Mexico State University, New York University, University of Notre Dame, Observat\'ario Nacional / MCTI, The Ohio State University, Pennsylvania State University, Shanghai Astronomical Observatory, United Kingdom Participation Group, Universidad Nacional Aut\'onoma de M\'exico, University of Arizona, University of Colorado Boulder, University of Oxford, University of Portsmouth, University of Utah, University of Virginia, University of Washington, University of Wisconsin, Vanderbilt University, and Yale University.

The DECam Local Volume Exploration Survey (DELVE; NOAO Proposal ID 2019A-0305, PI: Drlica-Wagner) is partially supported by Fermilab LDRD project L2019-011 and the NASA Fermi Guest Investigator Program Cycle 9 No. 91201.
This project used data obtained with the Dark Energy Camera (DECam), which was constructed by the Dark Energy Survey (DES) collaboration. Funding for the DES Projects has been provided by the U.S. Department of Energy, the U.S. National Science Foundation, the Ministry of Science and Education of Spain, the Science and Technology Facilities Council of the United Kingdom, the Higher Education Funding Council for England, the National Center for Supercomputing Applications at the University of Illinois at Urbana–Champaign, the Kavli Institute of Cosmological Physics at the University of Chicago, the Center for Cosmology and Astro-Particle Physics at the Ohio State University, the Mitchell Institute for Fundamental Physics and Astronomy at Texas A\&M University, Financiadora de Estudos e Projetos, Fundação Carlos Chagas Filho de Amparo à Pesquisa do Estado do Rio de Janeiro, Conselho Nacional de Desenvolvimento Científico e Tecnológico and the Ministério da Ciência, Tecnologia e Inovação, the Deutsche Forschungsgemeinschaft and the Collaborating Institutions in the Dark Energy Survey.
The Collaborating Institutions are Argonne National Laboratory, the University of California at Santa Cruz, the University of Cambridge, Centro de Investigaciones Enérgeticas, Medioambientales y Tecnológicas–Madrid, the University of Chicago, University College London, the DES-Brazil Consortium, the University of Edinburgh, the Eidgenössische Technische Hochschule (ETH) Zürich, Fermi National Accelerator Laboratory, the University of Illinois at Urbana-Champaign, the Institut de Ciències de l'Espai (IEEC/CSIC), the Institut de Física d'Altes Energies, Lawrence Berkeley National Laboratory, the Ludwig-Maximilians Universität München and the associated Excellence Cluster Universe, the University of Michigan, the National Optical Astronomy Observatory, the University of Nottingham, the Ohio State University, the OzDES Membership Consortium, the University of Pennsylvania, the University of Portsmouth, SLAC National Accelerator Laboratory, Stanford University, the University of Sussex, and Texas A\&M University.
Based in part on observations at Cerro Tololo Inter-American Observatory, National Optical Astronomy Observatory, which is operated by the Association of Universities for Research in Astronomy (AURA) under a cooperative agreement with the National Science Foundation.
Database access and other data services are hosted by the Astro Data Lab at the Community Science and Data Center (CSDC) of the National Science Foundation's National Optical Infrared Astronomy Research Laboratory, operated by the Association of Universities for Research in Astronomy (AURA) under a cooperative agreement with the National Science Foundation.

% \end{acknowledgments}

%% To help institutions obtain information on the effectiveness of their 
%% telescopes the AAS Journals has created a group of keywords for telescope 
%% facilities.
%
%% Following the acknowledgments section, use the following syntax and the
%% \facility{} or \facilities{} macros to list the keywords of facilities used 
%% in the research for the paper.  Each keyword is check against the master 
%% list during copy editing.  Individual instruments can be provided in 
%% parentheses, after the keyword, but they are not verified.

\vspace{5mm}
% \facilities{To be filled}
\facilities{CFHT (MegaCam), Gaia}

%% Similar to \facility{}, there is the optional \software command to allow 
%% authors a place to specify which programs were used during the creation of 
%% the manuscript. Authors should list each code and include either a
%% citation or url to the code inside ()s when available.

\software{\texttt{astropy} (\citealt{astropy2022}), \texttt{numpy} (\citealt{numpy2020}), Topcat (\citealt{topcat2005}), \texttt{gala} (\citealt{gala2017}), \texttt{scipy} (\citealt{scipy2020}), \texttt{emcee} (\citealt{foreman-mackey2013}), \texttt{matplotlib} (\citealt{matplotlib2007}), \texttt{AGAMA} (\citealt{vasiliev2018_agama})}

%% Appendix material should be preceded with a single \appendix command.
%% There should be a \section command for each appendix. Mark appendix
%% subsections with the same markup you use in the main body of the paper.

%% Each Appendix (indicated with \section) will be lettered A, B, C, etc.
%% The equation counter will reset when it encounters the \appendix
%% command and will number appendix equations (A1), (A2), etc. The
%% Figure and Table counter will not reset.

% \newpage
\appendix
\counterwithin{figure}{section} % Changes numbering to A.1, A.2, etc.

\section{Coordinate Transforms}
\label{chapter:appendix}

Equations from \citet{stoughton2002_SDSScoords} used to convert from SDSS survey coordinates to equatorial:

% \begin{equation}
%     \begin{split}
%         \cos(\alpha - 95\degree)\cos(\delta) = -\sin(\lambda_{SDSS}) \\
%         \sin(\alpha - 95\degree)\cos(\delta) = \cos(\lambda_{SDSS})\cos(\eta_{SDSS} + 32.5\degree) \\
%         \sin(\delta) = \cos(\lambda_{SDSS})\sin(\eta_{SDSS} + 32.5\degree)
%     \end{split}
% \end{equation}

\begin{align}
    \cos(\alpha - 95^{\circ})\cos(\delta) = -\sin(\lambda_{SDSS}) \\
    \sin(\alpha - 95^{\circ})\cos(\delta) = \cos(\lambda_{SDSS})\cos(\eta_{SDSS} + 32.5^{\circ}) \\
    \sin(\delta) = \cos(\lambda_{SDSS})\sin(\eta_{SDSS} + 32.5^{\circ})
\end{align}

\noindent where the origin ($\lambda$,~$\eta$)$_{SDSS}$ = (0$^{\circ}$,~0$^{\circ}$) and pole ($\lambda$,~$\eta$)$_{SDSS}$ = (0$^{\circ}$,~90$^{\circ}$) in SDSS survey coordinates are located at (RA,~Dec) = (185$^{\circ}$,~32.5$^{\circ}$) and (RA,~Dec) = (275$^{\circ}$,~0$^{\circ}$), respectively. The equatorial pole (RA,~Dec) = (0$^{\circ}$,~90$^{\circ}$) is located at ($\lambda$,~$\eta$)$_{SDSS}$ = (5.57$^{\circ}$,~0$^{\circ}$) in this frame of reference. 

% and pole in SDSS survey coordinates is

% is located at (RA, Dec) = (185$\degree$, 32.5$\degree$)

\section{Additional Figures}
\label{chapter:appendix_figs}

\begin{figure*}
    \centering
    \includegraphics[width=1\textwidth]{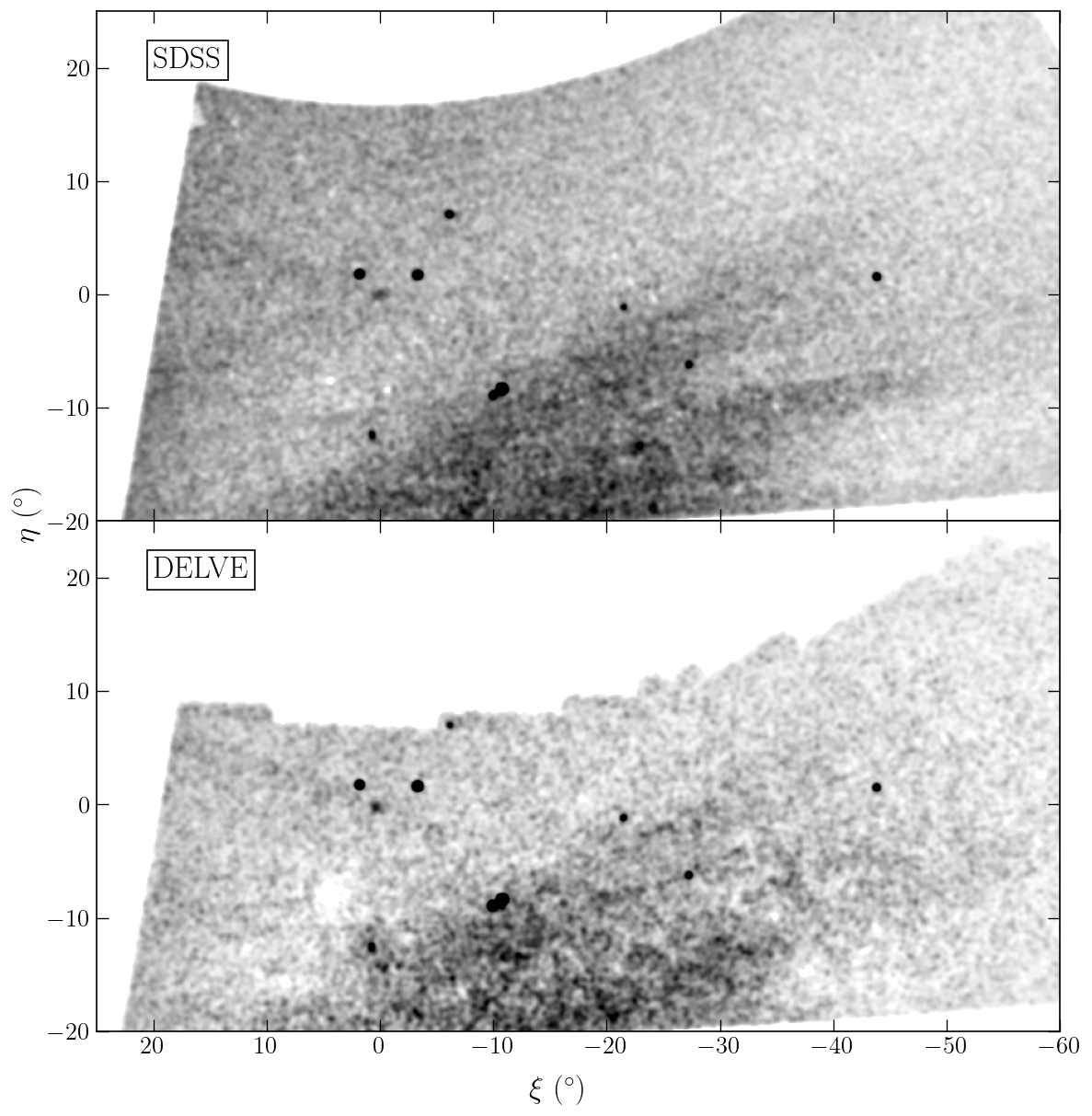}
    \caption{Matched filter maps of SDSS and DELVE, spanning a wider area to highlight the presence of the Sagittarius stellar stream. This feature is clearly observed in both maps as the overdensity ranging from $0^{\circ}~\lesssim~\xi~\lesssim~-40^{\circ}$.}
    \label{fig:MF_full}
\end{figure*}
% These two features are known as the stream's upper and lower bifurcations.

\begin{figure*}
    \centering
    \includegraphics[width=\textwidth]{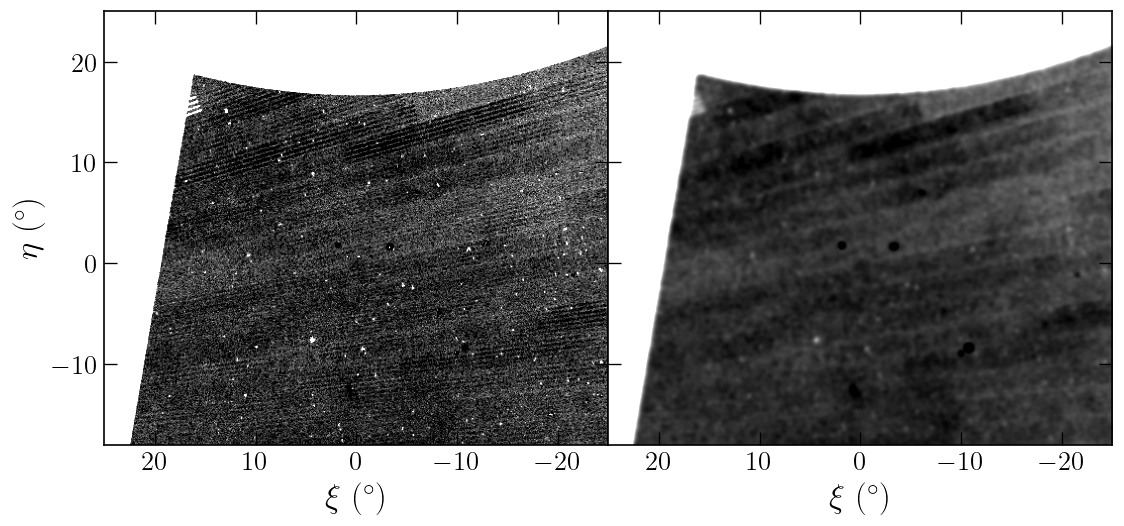}
    \caption{Tangent plane density maps of the matched filter likelihoods (left; pixel size of 0.1$^{\circ}$) and the same map smoothed with a Gaussian kernel of 0.2$^{\circ}$ (right) using the SDSS catalogue. The photometric limits of this data are only the 5-$\sigma$ point source depth determined in this work. As shown in both panels, the streaking observation pattern is oriented in the same direction as the faint feature seen in the SDSS panels of Figure \ref{fig:MF_map_tangent}. This feature is likely an artifact caused by the SDSS observing pattern that, when smoothed, appears to grow in size when summed together. As it is a feature that does not appear in DELVE MF maps, it is not likely a real MW substructure.}
    \label{fig:SDSS_streaking}
\end{figure*}

% \newpage

\begin{figure*}
    \centering
    \includegraphics[width=\textwidth]{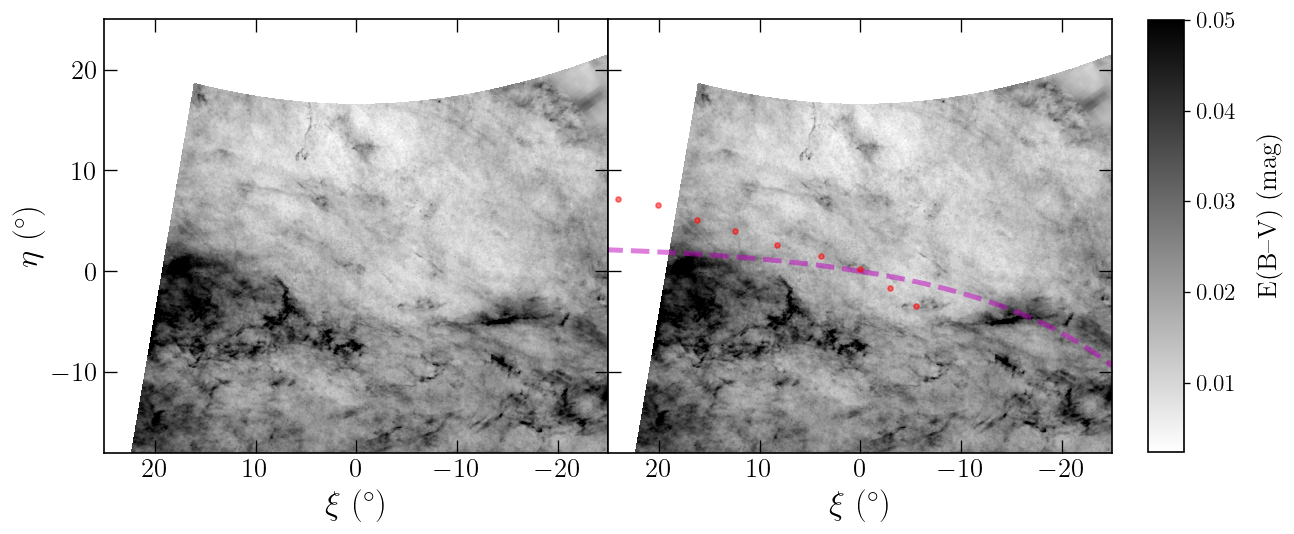}
    \caption{Extinction maps of the vicinity of Boo3, converted to tangent plane coordinates. Boo3's orbit and the approximate location of Styx are indicated in the magenta dashed line and red nodes, respectively, in the right panel for comparison.}
    \label{fig:extinction}
\end{figure*}

\bibliography{biblio}
\bibliographystyle{aasjournal}

%% This command is needed to show the entire author+affiliation list when
%% the collaboration and author truncation commands are used.  It has to
%% go at the end of the manuscript.
%\allauthors

%% Include this line if you are using the \added, \replaced, \deleted
%% commands to see a summary list of all changes at the end of the article.
%\listofchanges

\end{document}